\documentclass[12pt,preprint,preprint,nofootinbib]{revtex4}
\pdfoutput=1
\usepackage{epsf, array, color}
\usepackage{graphicx}
\usepackage{subfigure}
\usepackage{bbold}
\usepackage{amsmath}
\usepackage{amssymb}
\usepackage{mathtools}
\usepackage{epigraph}
\usepackage[utf8]{inputenc}
\usepackage{graphicx}
\usepackage[linktocpage,colorlinks,urlcolor=blue]{hyperref}
\usepackage{textcomp}

\usepackage{url}

\usepackage{dcolumn}

\begin{document}

\def\fb{\rm fb}
\def\etal{{\it et al.}}
\newcommand{\bea}{\begin{eqnarray}}
\newcommand{\eea}{\end{eqnarray}}
\newcommand{\nn}{\nonumber \\}
\newcommand{\lag}{\ensuremath{{\cal L}}} 
\newcommand{\Tr}{\operatorname{Tr}}
\newcommand{\Log}{\operatorname{Log}}
\newcommand{\Det}{\operatorname{Det}}
\def\beq{\begin{equation}}
\def\eeq{\end{equation}}
\newcommand{\vev}[1]{\langle {#1} \rangle}
\newcommand{\lsim}{\lesssim}
\newcommand{\ord}[1]{\mathcal{O}{(#1)}}
\newcommand{\co}{\mathcal{O}}
\newcommand{\gsim}{\gtrsim}
\newcommand{\eq}[1]{Eq.~(\ref{#1})}

\newcommand{\nc}{\newcommand}

\newcommand{\C}{\mathcal{C}}
\newcommand{\Op}{{\cal O}}
\nc{\vp}{\phi}
\nc{\tvp}{\widetilde{\phi}}
\nc{\vpj }{\mbox{${\vp^\dag i\,\raisebox{2mm}{\boldmath ${}^\leftrightarrow$}\hspace{-4mm} D_\mu\,\vp}$}}
\nc{\vpjt}{\mbox{${\vp^\dag i\,\raisebox{2mm}{\boldmath ${}^\leftrightarrow$}\hspace{-4mm} D_\mu^{\,a}\,\vp}$}}

\def\nn{\nonumber}
\def\gev{\rm GeV}
\def\tev{\rm TeV}
\def\mev{\rm MeV}
\def\ev{\rm eV}
\def\met{\rm MET}
\def\br{{\tt Br}}
\def\higgs{{\rm Higgs}}
\newcommand{\vLam}{{v^2\over\Lambda^2}}
\newcommand{\TphiWB}{{\C_{\phi WB}}}
\newcommand{\Tll}{\C_{ll}}
\newcommand{\Tlu}{\C_{lu}}
\newcommand{\Tld}{\C_{ld}}
\newcommand{\Tle}{\C_{le}}
\newcommand{\Ted}{\C_{ed}}
\newcommand{\Tqe}{\C_{qe}}
\newcommand{\Teu}{\C_{eu}}
\newcommand{\Tee}{\C_{ee}}
\newcommand{\Tdd}{\C_{dd}}
\newcommand{\Tuu}{\C_{uu}}
\newcommand{\Tfe}{\C_{\phi e}}
\newcommand{\Tfd}{\C_{\phi d}}
\newcommand{\Tfu}{\C_{\phi u}}
\newcommand{\TuB}{\C_{uB}}
\newcommand{\TuW}{\C_{uW}}
\newcommand{\TW}{\C_{W}}
\newcommand{\Tfla}{\C_{\phi l}^{(1)}}
\newcommand{\Tfqa}{\C_{\phi q}^{(1)}}
\newcommand{\Tfqc}{\C_{\phi q}^{(3)}}
\newcommand{\Tlqa}{\C_{lq}^{(1)}}
\newcommand{\Tlqc}{\C_{lq}^{(3)}}
\newcommand{\Tqqa}{\C_{qq}^{(1)}}
\newcommand{\Tqqc}{\C_{qq}^{(3)}}
\newcommand{\TphiD}{\C_{\phi D}}
\newcommand{\Tflc}{\C_{\phi l}^{(3)}}
\newcommand{\TphiB}{\C_{\phi B}}
\newcommand{\TphiW}{\C_{\phi W}}
\newcommand{\Tphik}{\C_{\phi \square}}
\newcommand{\Tuda}{\C_{ud}^{(1)}}
\newcommand{\Tqda}{\C_{qd}^{(1)}}
\newcommand{\Tqua}{\C_{qu}^{(1)}}

\preprint{IFT-UAM/CSIC-19-118}

\title{Electroweak  and QCD Corrections to $Z$  and $W$ pole observables in the SMEFT}

\author{Sally Dawson$^{\, a}$ and Pier Paolo Giardino$^{\, b}$ }
\affiliation{
\vspace*{.5cm}
  \mbox{$^a$Department of Physics,\\
  Brookhaven National Laboratory, Upton, N.Y., 11973,  U.S.A.}\\
 \mbox{$^b$ Instituto de Fisica Teorica, UAM-CSIC, Cantoblanco, 28049,
 Madrid, Spain}
 \vspace*{1cm}}

\date{\today}

\begin{abstract}
 We compute the next-to-leading order QCD and electroweak corrections to $Z$ and $W$ pole observables using the dimension-6 Standard Model effective
 field theory and present numerical results that can easily be included in global fitting programs.  Limits on SMEFT coefficient functions are presented at leading order and at next-to-leading order under several assumptions.   
\end{abstract}

\maketitle

\section{Introduction}

The LHC experiments provide strong evidence that the $SU(3)\times SU(2)\times U(1)$ Standard Model (SM) gauge theory describes physics at the electroweak scale\cite{Dawson:2018dcd}.  
To date, there is no evidence of new interactions or high mass particles.  Taken together, these features suggest that the 
weak scale can be described by an effective field theory (SMEFT) having the  SM as its low energy limit.  The SMEFT is defined by an infinite tower of on-shell and higher operators,  involving only the SM particles and assumes that the Higgs boson is part of an $SU(2)$ 
doublet\cite{Brivio:2017vri}. The effects of the higher dimension operators are suppressed by powers of a high scale, $\Lambda$, and we assume that  the most numerically relevant  operators are those of dimension-6.  All possible new physics phenomena are contained in the coefficient
functions. 

Numerous studies have been performed extracting limits on the coefficients of dimension-6 operators from 
global fits to Higgs measurements, vector boson pair production, electroweak measurements at the $Z$ and $W$ poles, top quark measurements, and low energy 
data\cite{Biekotter:2018rhp,Almeida:2018cld,deBlas:2017wmn,DiVita:2017eyz,Ellis:2018gqa,Grojean:2018dqj,Berthier:2016tkq,Pomarol:2013zra}.
 Typically, these fits use the most accurately known SM predictions, while the SMEFT effects are treated at lowest order (LO).  A program of calculations has begun to treat the SMEFT contributions at NLO, for both the QCD and electroweak (EW) contributions.  The SMEFT QCD corrections to gauge boson pair production\cite{Baglio:2017bfe,Baglio:2018bkm} and top quark production and decay\cite{Hartland:2019bjb,BuarqueFranzosi:2017jrj,Durieux:2018ggn,Boughezal:2019xpp} are known.  The electroweak SMEFT corrections to Higgs decays to
 $b\overline {b}$\cite{Cullen:2019nnr,Gauld:2016kuu,Gauld:2015lmb}, $\gamma\gamma$\cite{Dawson:2018liq,Dedes:2018seb,Hartmann:2015aia,Hartmann:2015oia}, $Z\gamma$\cite{Dawson:2018pyl,Dedes:2019bew}, $ZZ$\cite{Dawson:2018pyl}, and $WW$\cite{Dawson:2018liq} have also been computed, along with partial corrections to 
 the Drell Yan process\cite{Dawson:2019xfp}.  The SMEFT NLO corrections to $Z$ pole decays at NLO are also only partially known\cite{Dawson:2018jlg,Trott:2017yhn,Hartmann:2016pil}.

In this work, we take a major step  by computing the  next-to-leading order (NLO) 
EW and QCD corrections in the SMEFT to $Z$ and $W$  pole observables.  
 We assume flavor 
universality and use the Warsaw basis\cite{Grzadkowski:2010es}.    We are particularly interested in the numerical effects of the NLO corrections on the  global fits. In Section \ref{sec:basics}, we review the basics of the SMEFT theory and in Section \ref{sec:zpole}, we describe our NLO calculations.  Our results are given in Section \ref{sec:results} and Appendix \ref{sec:appresults}, which contains numerical 
expressions for the $Z$ and $W$ pole observables, as well as limits on the SMEFT coefficients at LO and NLO.  Section \ref{sec:conc}  contains
some conclusions and a discussion of the implications of our results for global fits.

\section{SMEFT basics}
\label{sec:basics}

The SMEFT
parameterizes new physics through an expansion in higher dimensional operators,
\begin{equation}
{\cal L}={\cal L}_{SM}+\Sigma_{k=5}^{\infty}\Sigma_{i=1}^n {\mathcal{C}_i^k\over \Lambda^{k-4}} O_i^k\, ,
\label{eq:lsmeft}
\end{equation}
where the $SU(3)\times SU(2)_L\times U(1)_Y$ invariant 
dimension-$k$ operators are constructed from SM fields and all of  the effects of the beyond the SM
 (BSM)  physics  reside in the coefficient functions, $\C_i^k$. 
 We assume all coefficients are real and do not consider the effects of CP violation.
We use the Warsaw basis \cite{Grzadkowski:2010es}
and at tree level (neglecting flavor)  there are $10$ dimension-$6$ operators contributing to the $Z$ and $W$ pole observables of
our study.  These operators are listed in Table \ref{tab:opdef}, where $\phi$ is the $SU(2)_L$ doublet,  $\tau^a$  are the Pauli matrices, 
$D_\mu=\partial_\mu +ig_s
T^AG_\mu^A +ig_2 {\tau^a\over 2}W_\mu^a+ig_1Y B_\mu$,
$q^T=(u_L,d_L)$,  $l^T=(\nu_L, e_L)$,   $W_{\mu\nu}^ a=
\partial_{\mu}W^a_{\nu} - \partial_{\nu}W^a_{\mu}
-g_2\epsilon^{abc}W^b_{\mu}W^c_{\nu}$, 
$\vpj=i\phi^\dagger(D_{\mu} \phi)
-i(D_{\mu}\phi)^\dagger \phi$,
and
$\vpjt=i\phi^\dagger \tau ^a D_{\mu} \phi
-i(D_{\mu}\phi)^\dagger \tau^a\phi$. 
 
\begin{table}[t] 
\centering
\renewcommand{\arraystretch}{1.5}
\begin{tabular}{||c|c||c|c||c|c||} 
\hline \hline
${\Op_{ll}}$                   & $(\bar l \gamma_\mu l)(\bar l \gamma^\mu l)$  &    
  ${\Op}_{\phi W B}$ 
 &$ (\vp^\dag \tau^a \vp)\, W^a_{\mu\nu} B^{\mu\nu}$  &
$\Op_{\vp D}$   & $\left(\vp^\dag D^\mu\vp\right)^* \left(\vp^\dag D_\mu\vp\right)$ 
\\
\hline 
   ${\Op}_{\phi e}$  &   $(\vpj) (\overline {e}_R\gamma^\mu e_R)$  & ${\Op}_{\phi u}$ & $(\vpj) (\overline {u}_R\gamma^\mu u_R)$ &
  ${\Op}_{\phi d}$
       & $(\vpj) (\overline {d}_R\gamma^\mu d_R)$
  \\ \hline 
             ${\Op}_{\phi q}^{(3)}$ & $(\vpjt)(\bar q \tau^a \gamma^\mu q)$  &${\Op}_{\phi q}^{(1)}$
      &$(\vpj)(\bar q \tau^a \gamma^\mu q)$  &
 ${\Op_{\vp l}^{(3)}}$      & $(\vpjt)(\bar l \tau^a \gamma^\mu l)$
 \\
\hline
${\Op}_{\phi l}^{(1)}$
      &  $(\vpj)(\bar l \tau^a \gamma^\mu l)$ &
    && &

\\
\hline \hline
\end{tabular}
\caption{Dimension-6 operators  contributing to the  $Z$ and $W$ pole observables of this study at tree level. \label{tab:opdef}}
\end{table}
At NLO, there are $22$ additional operators that  contribute:
\begin{eqnarray}&&
{\Op}_{ed}\, ,{\Op}_{ee}\, ,{\Op}_{eu}\, ,{\Op}_{lu}\,,{\Op}_{ld}\, ,{\Op}_{le}\, ,{\Op}_{lq}^{(1)}\, ,{\Op}_{lq}^{(3)}\, ,{\Op}_{\phi B}\, ,{\Op}_{\phi W}
\, ,{\Op}_{\square},
\nonumber \\
&&{\Op}_{qe}\, ,{\Op}_{uB}\, ,{\Op}_{uW}\, ,{\Op}_{W}\,, {\Op}_{qd}^{(1)}\,, {\Op}_{qq}^{(3)}\,,{\Op}_{qq}^{(1)}\,,{\Op}_{qu}^{(1)}\,,{\Op}_{ud}^{(1)}\,
,{\Op}_{uu}\,\,, {\Op}_{dd} \, .
\end{eqnarray}
Definitions for these operators can be found in Refs. \cite{Buchmuller:1985jz,Grzadkowski:2010es}.
We
 use the  Feynman rules  in $R_\xi$ gauge 
from Ref. \cite{Dedes:2017zog}.

The SMEFT interactions   cause the gauge field kinetic energies to have non-canonical normalizations and
following Ref.  \cite{Dedes:2017zog}, we define "barred" fields and couplings,
\begin{eqnarray}
{\overline W}_\mu^a & \equiv & (1-\C_{\phi W} v^2/\Lambda^2)W_\mu^a
\nonumber \\
{\overline B}_\mu & \equiv & (1-\C_{\phi B}v^2/\Lambda^2)B_\mu
\nonumber \\
{\overline g}_2 & \equiv &(1+\C_{\phi W} v^2/\Lambda^2)g_2
\nonumber \\
{\overline g}_1 & \equiv&  (1+\C_{\phi B}v^2/\Lambda^2)g_1\, ,
\end{eqnarray}
such  that ${\overline W}_\mu {\overline g}_2= W_\mu g_2$ and ${\overline B}_\mu {\overline g}_1= B_\mu g_1$. 
The "barred" fields have  canonically normalized  kinetic  energy interactions.
The masses of the W and Z fields to ${\cal {O}}\biggl({1\over \Lambda^2}\biggr)$ are
  \cite{Dedes:2017zog,Alonso:2013hga},
\bea
M_W^2&=&\frac{{\overline g}_2^2 v^2}4,\nn\\
M_Z^2&=&\frac{({\overline g}_1^2+{\overline g}_2^2) v^2}4+\frac{v^4}{\Lambda^2}\left(\frac18 ({\overline g}_1^2+{\overline g}_2^2) \C_{\phi D}+\frac12 {\overline g}_1{\overline g}_2\C_{\phi WB} \right).
\eea
Dimension-6 4-fermion operators    give contributions to the decay of the $\mu$, changing the relation between the 
vev, $v$, and the Fermi constant $G_\mu$, 
\begin{eqnarray}
G_\mu
\equiv \frac1{\sqrt{2} v^2}-\frac1{\sqrt{2}\Lambda^2}\C_{ll}+{\sqrt{2}\over \Lambda^2}\C_{\phi l}^{(3)}\, .
\label{eq:gdef}
\end{eqnarray}

The tree level SMEFT couplings of fermions  to the $Z$ and $W$ are given  in terms of our input parameters ($\alpha, M_Z, G_\mu$),
\begin{eqnarray}
  L&\equiv & 2M_Z \sqrt{\sqrt{2}G_\mu } Z_\mu\biggl[g_L^{Zq}+\delta g_{L}^{Zq}\biggr]
  {\overline q}\gamma_\mu q
 +  2M_Z \sqrt{\sqrt{2}G_\mu}   Z_\mu\biggl[g_R^{Zu}+\delta g_{R}^{Zu}\biggr]{\overline u}_R\gamma_\mu u_R
 \nonumber \\ &&
 +  2M_Z \sqrt{\sqrt{2}G_\mu}   Z_\mu\biggl[g_R^{Zd}+\delta g_{R}^{Zd}\biggr]{\overline d}_R\gamma_\mu d_R +
2M_Z \sqrt{\sqrt{2}G_\mu} Z_\mu\biggl[g_L^{Zl}+\delta g_{L}^{Zl}\biggr]
  {\overline l}\gamma_\mu l \nonumber \\ && 
 +  2M_Z \sqrt{\sqrt{2}G_\mu}   Z_\mu\biggl[g_R^{Ze}+\delta g_{R}^{Ze}\biggr]
  {\overline e}_R\gamma_\mu e_R
  +  2M_Z \sqrt{\sqrt{2}G_\mu}   \biggl( \delta g_{R}^{Z\nu}\biggr)
  {\overline \nu}_R\gamma_\mu \nu_R
  \nonumber \\
  &&+{{\overline{g}}_2\over \sqrt{2}}\biggl\{W_\mu\biggl[(1+\delta g_{L}^{Wq}){\overline u}_L\gamma_\mu d_L
  +\biggl(\delta g_R^{Wq}\biggr) 
  {\overline u}_R\gamma_\mu d_R\biggr] 
  \nonumber \\
  &&+W_\mu\biggl[(1+\delta g_{L}^{Wl}){\overline \nu}_L\gamma_\mu e_L
  +\biggl( \delta g_R^{W\nu}\biggr)
  {\overline \nu}_R\gamma_\mu e_R\biggr] 
  +h.c.\biggr\}\, .
  \label{eq:dgdef}
  \end{eqnarray}

  We assume  all couplings are flavor independent and we neglect
 CKM mixing.  The weak coupling in Eq. \ref{eq:dgdef} is evaluated using the LO SM relation and Eq \ref{eq:sw}
 serves as the definition of $s_W^2$,
 \begin{eqnarray}
 {\overline{g}}_2^2&=&2\sqrt{2}G_\mu M_Z^2\biggl(1+\sqrt{ 1-{4\pi\alpha\over\sqrt{2}G_\mu M_Z^2}}\biggr)\, . 
 \nonumber \\
 s_W^2&\equiv & {4\pi\alpha\over {\overline{g}}_2^2}\, .
 \label{eq:sw}
 \end{eqnarray}
 Since we are working to ${\cal{O}}\biggl({v^2\over \Lambda^2}\biggr)$,  we omit
 dipole type operators that do not interfere with the SM contributions to $Z$ and $W$ pole observables.  Similarly,
 the contributions from right-handed $W$ couplings and the right-handed $Z {\overline{\nu}}\nu$  interaction do not contribute to our study. 
 The tree level couplings are,
\begin{eqnarray}
g_R^{Zf}&=&-s_W^2 Q_f\quad{\rm and}\quad g_L^{Zf}=T_3^f -s_W^2 Q_f
\end{eqnarray}
with $T_3^f=\pm \displaystyle \frac{1}{2}$.   $SU(2)$ invariance implies,
\begin{eqnarray}
\delta g_L^{Wq}&=&\delta g_L^{Zu}-\delta g_L^{Zd}\nonumber \\
\delta g_L^{Wl}&=&\delta g_L^{Z\nu}-\delta g_L^{Ze}\, . 
\label{eq:su2rel}
\end{eqnarray}
The SMEFT contributions to the effective couplings are listed in Table \ref{tab:ferm}\cite{Berthier:2015oma}.

\begin{table}
\centering
\begin{tabular}{|c||c|}
\hline
& Warsaw  Basis \\
\hline\hline
$\delta g_L^{Zu}$&$ -\frac{v^2}{2\Lambda^2}\left(\C_{\phi q}^{(1)}-\C_{\phi q}^{(3)}\right) + \frac12 \delta g_Z + \frac23\left(\delta s_W^2 - s_W^2 \delta g_Z^{}\right)$ \\
\hline
$\delta g_L^{Zd}$& $-\frac{v^2}{2\Lambda^2}\left(\C_{\phi q}^{(1)}+\C_{\phi q}^{(3)}\right) - \frac12 \delta g_Z - \frac13\left(\delta s_W^2 - s_W^2 \delta g_Z^{}\right)$\\
\hline
$\delta g_L^{Z\nu}$& 
$ -\frac{v^2}{2\Lambda^2}\left(\C_{\phi l}^{(1)}-\C_{\phi l}^{(3)}\right) + \frac12 \delta g_Z $
\\
\hline
$\delta g_L^{Ze}$&  $-\frac{v^2}{2\Lambda^2}\left(\C_{\phi l}^{(1)}+\C_{\phi l}^{(3)}\right) - \frac12 \delta g_Z - \left(\delta s_W^2 - s_W^2 \delta g_Z^{}\right)$\\
\hline
$\delta g_R^{Zu}$&$-\frac{v^2}{2\Lambda^2} \C_{\phi u} + \frac23\left(\delta s_W^2 - s_W^2\delta g_Z^{}\right)$ \\
\hline
$\delta g_R^{Zd}$& $-\frac{v^2}{2\Lambda^2} \C_{\phi d} - \frac13\left(\delta s_W^2 - s_W^2\delta g_Z^{}\right)$\\
\hline
$\delta g_R^{Ze}$& $-\frac{v^2}{2\Lambda^2} \C_{\phi e} - \left(\delta s_W^2 - s_W^2\delta g_Z^{}\right)$ \\
\hline
$\delta g_L^{Wq}$&  $ \frac{v^2}{\Lambda^2}\C_{\phi q}^{(3)} + c_W^2\delta g_Z^{} + \delta s_W^2$\\
\hline
$\delta g_L^{W l}$&   $ \frac{v^2}{\Lambda^2}\C_{\phi l}^{(3)} + c_W^2\delta g_Z^{} + \delta s_W^2$\\
\hline
$\delta g_Z$ & $-\frac{v^2}{\Lambda^2}\left(\delta v +\frac14\C_{\phi D}^{}\right)$ \\ 
\hline
$\delta v$  &$\C_{\phi l}^{(3)} - \frac12 \C_{ll}^{}$ \\
\hline
$\delta s_W^2$  & $ -\frac{v^2}{\Lambda^2} \frac{s_W^{} c_W^{}}{c_W^2-s_W^2}\left[2 s_W^{} c_W^{}\left(\delta v + \frac14 \C_{\phi D}^{}\right) + \C_{\phi WB}^{}\right]$ \\
\hline
\hline
\end{tabular}
\caption{Anomalous fermion couplings at LO in the Warsaw~\cite{Grzadkowski:2010es} basis. }
\label{tab:ferm}
\end{table}

\section{$W$ and $Z$ pole observables to NLO}
\label{sec:zpole}

The observables we consider are:
\begin{eqnarray}
&&M_W, \Gamma_W, \Gamma_Z,  \sigma_h,  R_l, A_{l,FB}, R_b, R_c,  A_{FB,b}, A_{FB,c},  A_b, A_c, A_l\, .
\label{eq:quan}
\end{eqnarray} 
The SM results for these observables are quite precisely known, and as a by product of our study we recover the known
NLO QCD and NLO EW results as a check of our calculation\cite{Hollik:1988ii}.

\begin{figure}
  \centering
\includegraphics[width=1\textwidth]{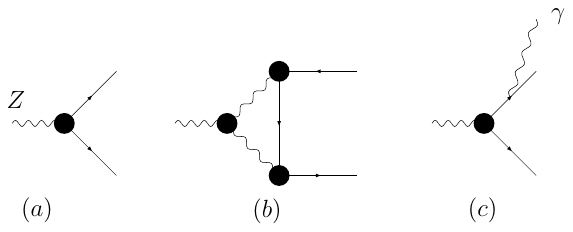}
\vskip -5in
 \caption{ {Sample electroweak diagrams contributing to $Z\rightarrow f {\overline {f}}$ at NLO in the SMEFT: (a) Tree level SMEFT diagram,
 (b) virtual SMEFT diagram, and (c) real photon emission in the SMEFT.  The circles represent potential insertions of
 dimension-6 SMEFT operators.
   \label{fig:diag}}}
\end{figure}

The next-to-leading order contributions to $Z$ and $W$ pole observables require the calculation of  one
loop virtual diagrams in the SMEFT and in most cases,  the  contribution also of real photon  and gluon emission diagrams.
Since the SMEFT theory is renormalizable order by order in the $(v^2/\Lambda^2)$ expansion, 
we retain only terms of  ${\cal{O}}(v^2/\Lambda^2)$.
The one-loop SMEFT calculations contain  both tree level
and one-loop
contributions from the dimension-6 operators, along with the  full electroweak and QCD one-loop SM amplitudes.
Sample diagrams contributing to the $Z$ decay widths at NLO are shown in Fig. \ref{fig:diag}.  
For $W$ decays, there are also
 dipole-like $\gamma (g)W f{\overline f}$ contact interactions that we  include.  (The corresponding $\gamma (g)Z f{\overline f}$ 
 operators first arise at dimension-8.)
Since we concentrate on the $Z$ pole physics, we calculate the cross sections,  $e^+e^-\rightarrow$  hadrons,
using the narrow width approximation:
\begin{equation}
\sigma^0_{\rm had}=\sum_{f=u,d,s,c,b}\frac{12\pi}{M_Z^2}\frac{\Gamma_e\Gamma_f}{\Gamma_Z^2}.
\end{equation}
Corrections to this formula are of higher order and we do not include them \cite{Dubovyk:2019szj}. 
Non-resonant contributions, such as photon exchange, box diagrams and 4-fermions interactions \cite{Berthier:2015oma}, are also not included
because they do not contribute to the observables on the $Z$ pole to ${\cal{O}}({1\over\Lambda^2})$.

We employ a modified on-shell (OS) scheme, where the SM parameters are renormalized in the OS scheme. 
The  effective field theory coefficients of the dimension-6 operators are treated as $\overline{{MS}}$
parameters and the poles of the  one-loop coefficients $\C_i$ are known from Refs.  \cite{Jenkins:2013zja,Jenkins:2013wua,Alonso:2013hga},
\beq
\C_i(\mu)=\C_{0,i}-\frac1{2\hat{\epsilon}}\frac1{16\pi^2}\gamma_{ij}\C_j,
\eeq
where $\mu$ is the renormalization scale,  $\gamma_{ij}$ is the one-loop anomalous dimension, 
\beq
\mu \frac{d \C_i}{d\mu}=\frac1{16\pi^2}\gamma_{ij}\C_j,
\eeq
and $\hat{\epsilon}^{-1}\equiv\epsilon^{-1}-\gamma_E+\log(4\pi)$.

The renormalized SM  gauge boson masses are,
\beq
M_V^2=M^2_{0,V}-\Pi_{VV}(M^2_{V}),
\eeq
where $\Pi_{VV}(M^2_V)$ is the one-loop correction to the 2-point function for Z or W computed on-shell and tree level quantities
are denoted with the subscript $0$ in this section. 
The gauge boson $2$-point functions in the SMEFT can be found  analytically in Refs.  \cite{Chen:2013kfa,Ghezzi:2015vva}.
The one-loop relation between the vacuum expectation value and the Fermi constant is,
\beq
G_\mu+{\C_{ll}\over\sqrt{2}\Lambda^2}-\sqrt{2}{\C_{\phi l}^{(3)}\over\Lambda^2}\equiv\frac1{\sqrt{2} v_0^2}(1+\Delta r),
\label{eq:geftdef}
\eeq
where $v_0$ is the unrenormalized minimum of the potential and
 $\Delta r$ is obtained from the one-loop corrections to $\mu$ decay.
Complete  analytic expressions for $\Delta r$ in both the SM and the SMEFT at dimension-$6$ are given in Ref.  \cite{Dawson:2018pyl}.   Finally, the on-shell renormalization of $\alpha$ is extracted from the renormalization of the ${\overline l}l \gamma$ vertex.

 We obtain the relevant amplitudes  for the virtual contributions
 using FeynArts  \cite{Hahn:2000kx} with a model file generated by FeynRules \cite{Alloul:2013bka} and  the Feynman rules 
 of Ref.  \cite{Dedes:2017zog}. Then we use FeynCalc \cite{Mertig:1990an,Shtabovenko:2016sxi} to manipulate and reduce the integrals and LoopTools \cite{Hahn:2000jm} for the numerical evaluation.

The $Z$ decays to charged fermions receive contributions from one-loop virtual diagrams and  from real photon emission that are separately IR divergent and
we regulate these divergences with a photon mass.  Since we only consider
the
 inclusive quantities of Eq. \ref{eq:quan}, the photon mass dependence cancels after integration over
the photon phase space and there is no need for a photon energy cut.
 The complicated form of the SMEFT vertices makes direct integration of the phase space difficult, so
 we use the method of Ref.   \cite{Anastasiou:2002yz}, where the integration over the photon phase space is replaced with a loop integration.   
This is possible after we use the identity, 
\bea
2i\pi\delta(p^2-m^2)=\frac1{p^2-m^2+i0}-\frac1{p^2-m^2-i0}.
\label{eq:Cutkosky}
\eea
After making this replacement, we treat the momenta of the outgoing particles as internal loop momenta,  the integration over the phase space becomes an integration over the loop momenta and we can use the IBP relations  to reduce the  loop integrals to known master integrals.  
In the case of $Z \to f {\overline{f}}\gamma$, the integrals are 2-point 2-loop integrals, for which a generic basis of master integrals is known  \cite{Tarasov:1997kx, Martin:2003qz}
and the reduction can be done using FIRE \cite{Smirnov:2014hma}.   This is identical to the technique we applied in the calculation of the real contributions to 
$H\rightarrow W^+W^-\gamma $ in Ref. \cite{Dawson:2018liq}.

\section{Results}
\label{sec:results}
We take as our  physical input parameters,
 \begin{eqnarray}
G_\mu&=&1.1663787(6)\times 10^{-5} \gev^{-2}\nonumber \\
M_Z&=&91.1876\pm .0021\gev\nonumber \\
{1\over \alpha} &=& {137.035999139(31)} \nonumber\\
\Delta\alpha_{\rm had}^{(5)} &=& 0.02764\pm 0.00009\nonumber\\
\alpha_s(M_Z)&=&0.1181\pm 0.0011 \nonumber \\
M_H&=&125.10\pm 0.14 ~\gev\nonumber\\
M_t&=&172.9\pm0.5~\gev\nonumber\, .
\label{eq:inputs}
\end{eqnarray}

The lowest order SMEFT contributions to Z pole observables, ${\cal {O}}_i$ are well known.  We write
the  SMEFT predictions for the observables as,
\begin{eqnarray}
O_i^{SMEFT, LO}&=&O_i^{SM}+\delta O_i^{LO}\nonumber \\
O_i^{SMEFT,NLO}&=&O_i^{SM}+\delta O_i^{NLO}\, ,
\label{eq:defs}
\end{eqnarray}
and we  present our results  numerically.
In Table  \ref{tab:expnums}, we summarize the current state of the  SM theory and the experimental results.  The
theory errors include the parametric uncertainties on $M_t$ and $M_H$\cite{Dubovyk:2019szj}.     
In evaluating $O_i^{SM}$ in Eq. \ref{eq:defs}, we always use the most accurately calculated value given in Table \ref{tab:expnums}.

\begin{table}
\begin{center}
\begin{tabular}{|l|c|c|}
\hline
Measurement& Experiment& "Best" theory
\\
\hline\hline
$\Gamma_Z$(GeV) & $2.4952\pm 0.0023$ & $2.4945\pm 0.0006$   \cite{Freitas:2014hra,Dubovyk:2018rlg,Dubovyk:2019szj}  
\\
\hline
$\sigma_h$(nb) & $41.540\pm 0.037$ &  $ 41.491\pm 0.008$\cite{Freitas:2014hra,Dubovyk:2018rlg,Dubovyk:2019szj}  \\
\hline
$R_l$ & $20.767\pm 0.025$\cite{ALEPH:2005ab} & $ 20.749\pm 0.009$\cite{Freitas:2014hra,Dubovyk:2018rlg,Dubovyk:2019szj}   \\
\hline 
$R_b$& $0.21629\pm 0.00066$ & $ 0.21586\pm 0.0001$\cite{Freitas:2014hra,Dubovyk:2018rlg,Dubovyk:2019szj}   \\
\hline
$R_c$ & $0.1721\pm 0.0030$ & $ 0.17221\pm 0.00005$\cite{Freitas:2014hra,Dubovyk:2018rlg,Dubovyk:2019szj}  
\\
\hline 
$A_l$ & $0.1465\pm 0.0033$\cite{ALEPH:2005ab} & $ 0.1472\pm 0.0004$ \cite{Dubovyk:2019szj,Awramik:2006uz}\\
\hline
$A_c$ & $0.670\pm 0.027$ & $ 0.6679\pm 0.0002$\cite{Dubovyk:2019szj,Awramik:2006uz}\\
\hline 
$A_b$ & $0.923\pm 0.020$ & $0.92699\pm 0.00006$\cite{Dubovyk:2019szj,Awramik:2006uz,Awramik:2008gi}
\\ \hline  
$A_{l,FB}$ & $0.0171\pm 0.0010$ & $ 0.0162\pm 0.0001$ \cite{Dubovyk:2019szj,Awramik:2006uz} \\
 \hline
$A_{b,FB}$ & $0.0992\pm 0.0016$ & $ 0.1023\pm 0.0003$ \cite{Dubovyk:2019szj,Awramik:2006uz,Awramik:2008gi} \\
\hline
$A_{c,FB}$ & $0.0707\pm 0.0035$ & $ 0.0737\pm 0.0003$  \cite{Dubovyk:2019szj,Awramik:2006uz}\\
\hline
$A_l(SLD)$ & $0.1513\pm 0.0021$\cite{ALEPH:2005ab} & $ 0.1472\pm 0.0004$\cite{Dubovyk:2019szj,Awramik:2006uz}\\
\hline
$\sin^2\theta_{l,eff}$ &$0.23179\pm 0.00035$ \cite{CDF:2016cry}& $ 0.23150\pm 0.00006$\cite{Dubovyk:2019szj,Awramik:2006uz,Awramik:2008gi}\\ \hline
$M_W $(GeV) & $80.379\pm 0.012$ \cite{PhysRevD.98.030001}&$ 80.359\pm 0.006$\cite{Awramik:2003rn,Erler:2019hds}\\ \hline
$\Gamma_W$(GeV)  & $2.085\pm 0.042$ \cite{PhysRevD.98.030001}& $ 2.0904\pm  0.0003$\cite{Cho:2011rk}\\
\hline
\end{tabular}
\caption{Experimental results and SM predictions for $W$ and $Z$ pole observables, assuming lepton universality 
The theory includes the full set of $2$-loop contributions
for the $Z$ pole observables, along with higher order corrections when known.   When not specified, the numbers 
are taken from Table 10.5 of  the electroweak review of Ref. \cite{PhysRevD.98.030001}.   The theory predictions are
computed using the formulae in the indicated references and our input parameters, and the theory errors include
the parametric uncertainties on $M_t$ and $M_H$\cite{Dubovyk:2019szj}, along with the estimated theory 
uncertainties described in the respective papers.\label{tab:expnums}}
\end{center}
\end{table}

We do not include
the effective weak leptonic mixing angle in our fit since it can be directly derived
from other observables, but present it here for completeness.  
\begin{eqnarray} 
\delta\sin^2\theta_{l,eff}^{LO} &=& \vLam\bigg\{ -0.28785 \Tfe - 0.21215\Tfla +0.36851\Tflc   -0.29033\Tll \nonumber \\ && + 
 0.14517 \TphiD + 0.71015\TphiWB
 \biggr\}
\nonumber \\ 
\delta\sin^2\theta_{l,eff}^{NLO} &=&\vLam\biggl\{- 0.2726 \Tfe - 0.23666 \Tfla + 0.42246 \Tflc  - 0.31904 \Tll \nonumber \\ &&
+ 0.16629 \TphiD + 0.77518 \TphiWB  \nonumber \\ &&
-0.00036 \Ted - 0.00100 \Tee + 0.00677 \Teu + 0.00161 \Tfd  + 0.01033 \Tfqa \nonumber \\ &&- 0.00871 \Tfqc  - 
 0.01424 \Tfu - 0.00028 \Tld - 0.00064 \Tle - 0.00401 \Tlqa  \nonumber \\ && - 0.00106 \Tlqc + 0.00531 \Tlu + 0.00032 \TphiB   + 0.00004 \Tphik + 0.00032 \TphiW \nonumber \\ && - 0.00512 \Tqe + 0.01087 \TuB + 0.00917 \TuW + 0.00053 \TW
 \biggr\}\, .
\nonumber \\
\end{eqnarray}
The NLO corrections  to $\sin^2\theta_{l,eff}$ change
the numerical effects of the coefficients appearing at tree level by ${\cal{O}}(5-10\%)$, and introduce dependencies on other coefficients. 

For the $W$ mass and  total width, we find the predictions,
\begin{eqnarray}
\delta M_W^{LO} &=&  \vLam\biggl\{  -29.827 \Tflc + 14.914 \Tll - 27.691 \TphiD - 57.479 \TphiWB
  \biggr\}\, \gev
\nonumber \\
\
\delta M_W^{NLO} &=&\vLam\biggl\{  - 35.666 \Tflc+ 17.243 \Tll - 30.272 \TphiD - 64.019 \TphiWB \nonumber \\ &&
-0.137 \Tfd - 0.137 \Tfe - 0.166 \Tfla - 2.032 \Tfqa + 
 1.409 \Tfqc  + 2.684 \Tfu  \nonumber \\ && + 0.438 \Tlqc - 0.027 \TphiB - 0.033 \Tphik - 0.035 \TphiW - 
 0.902 \TuB - 0.239 \TuW - 0.15 \TW
  \biggr\}\, \gev
  \nonumber \\ 
  \delta \Gamma_W^{LO}&=&\vLam
\biggl\{
 -5.092 \Tflc + 2.784 \Tfqc + 3.242 \Tll - 2.143 \TphiD - 4.448 \TphiWB
  \biggr\} \, \gev
\nonumber \\ 
\delta \Gamma_W^{NLO}&=&
\vLam\biggl\{- 5.556 \Tflc + 2.996 \Tfqc + 3.340 \Tll - 2.260 \TphiD - 4.777 \TphiWB \nonumber \\ &&
-0.01 \Tfd - 0.01 \Tfe - 0.017 \Tfla  - 0.153 \Tfqa  + 0.203 \Tfu+ 0.048 \Tlqc\nonumber \\ && - 0.002 \TphiB  - 0.003 \Tphik - 0.004 \TphiW  - 0.03 \Tqqa - 
 0.094 \Tqqc \nonumber \\ &&- 0.068 \TuB - 0.014 \TuW - 0.013 \TW
  \biggr\}\, \gev\, .
\end{eqnarray}
It is interesting to note that  some of the contributions to the $W$ mass  and width change by more than $10\%$ when going from LO to NLO in the SMEFT.
The NLO SMEFT contributions to  the other observables of Eq. \ref{eq:quan} given in Appendix \ref{sec:results}.

We fit to the experimental data given in Table \ref{tab:expnums}, (omitting $\sin^2 \theta_{eff}$) since it can be directly 
derived from other observables).  The most accurate SM predictions are given in the right-hand column 
and we use these values in our fits, as opposed to the LO or NLO SM contributions directly calculated. 
The pole observables we consider are\cite{Berthier:2015gja,Corbett:2017qgl,Falkowski:2014tna}:
\begin{eqnarray}
&&M_W, \Gamma_{W},  \Gamma_{Z},  \sigma_h,  R_l, A_{l,FB}, R_b, R_c,  A_{FB,b}, A_{FB,c},  A_b, A_c, A_l\, ,
\end{eqnarray} 
where we assume lepton universality and the experimental correlations can be found in Ref. \cite{ALEPH:2005ab}.  We include the
measurements of $A_l$ from LEP and SLD as separate data points. 

The $\chi^2$ is computed from, 
\begin{eqnarray}
\chi^2&=&\Sigma_{i,j}(O_i^{exp}-O_i^{SMEFT})\sigma^{-2}_{ij} (O_j^{exp}-O_j^{SMEFT}) \, .
\end{eqnarray}
Using the LO SMEFT expressions for the observables  and taking $\Lambda=1~TeV$, we find\footnote{Our results  are consistent with SMEFT fits to  purely LEP observables using slightly different
sets of inputs\cite{Falkowski:2014tna,Berthier:2016tkq,Corbett:2017qgl,Berthier:2015gja}. },
\begin{eqnarray}
\chi^2_{LO} &=&\chi^2_{SM}+
32 \Tfd + 105 \Tfe - 445 \Tfla \nonumber \\ && + 639 \Tflc - 
 49 \Tfqa - 60 \Tfqc - 11 \Tfu \nonumber \\ && - 424 \Tll + 
 491 \TphiD + 1114 \TphiWB +{\vec C}^T_{LO} M_{LO}{\vec C}_{LO}\, 
\end{eqnarray}
where 
\begin{eqnarray}
&&{\vec C}^T_{LO} =\biggl(\Tll,\, \TphiWB, \,\Tfu,\,
\Tfqc,\, \Tfqa,\,
 \Tflc,\,  \Tfla, \,  \Tfe, \, \TphiD, \,
\Tfd \biggr)
\end{eqnarray}
and we  find $\chi^2_{SM}\sim 13.42$.   The  symmetric matrix $M_{LO}$ is,
\begin{equation}
M_{LO}=\left(\begin{array}{rrrrrrrrrr}
25279  &-108322 & 1799  &14960  &4513  &-71171 & 27975  &
16835  & -37889  & -831 \\
&148456 &-851 &-11405 & -3882 & 165479 &-51962  &-55619   &102746  & -629 \\
 &   &574   & 6873   &  1615  & -7314 & 
-6662 & 3620   & -899  &  -697  \\
   &  &    & 24474 & 
13826   & -54867   &  -45834  &  27540   &- 7486 & -5161 \\
 &   &    &    &  3097  &
-15840  & -12754   &  8236   & -2257  & -1569    \\
&    &    & 
   &  & 70369  &  18870  & -56402 &   60803  & 4835 \\
   &  &   &  &    &    & 58121  & -44390   &-13987  & 5382  \\
 &      &   &  &  &  &  & 31734   & -8417 & -2193  \\
   &  & &    &  &   &    &  
 & 21176  
&  415 \\
&&&&&&&&
&  318  
\end{array}
\right)  \, .
\end{equation} 

Using the NLO SMEFT expressions we find $\chi^2_{NLO}$ , (for $\Lambda=1~TeV$) ,
\begin{eqnarray}
\chi^2_{NLO} &=&\chi^2_{SM}
-403 \Tll
+1070\TphiWB   
 -53\Tfu
  -93 \Tfqc
     \nonumber \\ &&
    -18 \Tfqa
    +666\Tflc     
   -402   \Tfla
   +176   \Tfe 
    +502 \TphiD 
   + 27   \Tfd 
  \nonumber \\ &&
   -1.48\Tqqa
  + 0.55\Tphik
  +0.62\TphiW 
   +0.48 \TphiB 
     +6.55\TuW
 \nonumber \\ &&
 +15 \TuB
  +0.23 \Ted
  +0.063\Tdd
+0.56 \Tee
  +1.40  \Tqqc +
 2.38\TW
 \nonumber \\ &&
+0.53 \Tuu
 -0.54 \Tuda
 +1.05 \Tqua
  -4.88   \Tlqc
+2.8 \Tqe
 +0.34 \Tqda
\nonumber \\ &&
+9.8 \Tlu
-0.32\Tle
-0.49 \Tld
 -3.8  \Teu-7.5 \C_{lq}^{(1)}
+{\vec C}^T_{NLO} M_{NLO}{\vec C}_{NLO}\, ,
\end{eqnarray}
where,
\begin{eqnarray}
{\vec C}^T_{NLO} &=&\biggl(\Tll,\, \TphiWB, \,\Tfu,\,
\Tfqc,\, \Tfqa,\,
 \Tflc,\,  \Tfla, \,  \Tfe, \, \TphiD, \,
\Tfd ,
\nonumber \\
&&\Ted\, ,\Tee\, ,\Teu\, ,\Tlu\,,\Tld\, ,\Tle\, ,\Tlqa\, ,\Tlqc\, ,\TphiB\, ,\TphiW
\, ,\Tphik,\nonumber \\
&&\Tqe\, ,\TuB\, ,\TuW\, ,\TW\,, \Tqda\,, \Tqqc\,,
\Tqqa\,,\Tqua\,,\Tuda\,
,\Tuu\,\,, \Tdd \, \biggr)\,  ,
\end{eqnarray}
where the numerical form of $M_{NLO}$ is given in the supplemental material.
At NLO, the $\chi^2$ now depends on $32$ coefficients, and the effects of the coefficients appearing at LO have shifted by $5-10\%$.
The (relatively) large shift of the coefficients of $\C_{\phi u}$ and $\C_{\phi Q}^{(1,3)}$ are  due to the top quark loop.

To study the numerical importance of the NLO effects, we begin by keeping only one coefficient non-zero at a time.
We find the $95\%$ confidence level regions at LO and NLO shown in Tables \ref{tab:lopred} and \ref{tab:nlopred}.   The
largest effect of the NLO corrections is on the coefficient of $\C_{\phi u}$.  The relatively large allowed values for $\C_{\phi d}$ are the
result of the discrepancy in  the measured value $A_{FB,b}$ from the  SM prediction.

\begin{table}
\begin{center}
\begin{tabular}{|c|c|c|}
\hline
Coefficient& LO & NLO\nonumber \\ 
\hline\hline
$\C_{ll}$ & $[ -0.0039,    0.021]$& $ [-0.0045,    0.019]$\\
\hline
$\C_{\phi WB}$& $[  -0.0088,    0.0013]$& $[ -0.0080,    0.0016]$\\
\hline
$\C_{\phi u}$ & $[ -0.072,    0.091]$ & $[ -0.035,    0.085]$\\
\hline
$\C_{\phi q}^{(3)}$ &  $[-0.011,    0.014]$ &  $[-0.010,    0.014]$\\
\hline
$\C_{\phi q}^{(1)}$  &  $[-0.027,    0.043]$& $[ -0.031,    0.036]$\\
\hline 
$\C_{\phi l}^{(3)}$ &  $[-0.012,    0.0029]$ &  $[ -0.010,    0.0028]$\\
\hline
$\C_{\phi  l}^{(1)}$ &  $[ -0.0043,    0.012]$ &  $[ -0.0047,    0.012]$\\
\hline
$\C_{\phi e}$ & $[ -0.013,    0.0094]$ & $ [-0.013,    0.0080]$\\
\hline
$\C_{\phi D}$ & $[ -0.025,    0.0019]$ & $[ -0.023,    0.0023]$\\
\hline
$\C_{\phi d}$ & $[ -0.16,        0.060]$ & $[-0.13,        0.064]$\\
\hline\hline
\end{tabular}
\caption{$95\%$ confidence level allowed ranges for single parameter fit to coefficients contributing to the
lowest order predictions. The scale $\Lambda$ is taken to be $1~TeV$. \label{tab:lopred}}
\end{center}
\end{table}
\begin{table}
\begin{center}
\begin{tabular}{|c|c||c|c||c|c|}
\hline
Coefficient& NLO & Coefficient & NLO & Coefficient  &NLO\nonumber \\ 
\hline\hline
$\C_W$ &  $[  -4.8,        0.48]$       & $\C_{uu}$ &  $[-1.1,        0.99$] & $\C_{uW}$ & $[-0.78,       0.29]$   \\
\hline
$\C_{uB}$ & $[ -0.57,        0.11]$  & $\C_{qu}^{(1)}$ &  $[-2.2,         1.3]$ &     $ \C_{qq}^{(3)}$ & $[-0.32,        0.29]$  \\
\hline
$\C_{qq}^{(1)}$ &  $[ -0.93,         1.5]$    & $\C_{qe}$ & $[-0.75,        0.48]$    &  $\C_{qd}^{(1)}$ &$[-9.8,         5.0]$\\   
\hline
$\C_{\phi \square}$ & $[ -22,  1.9]$   & $\C_{\phi W}$ & $ [ -17,  2.2]$   & $\C_{\phi B}$ & $[ -19   , 3.3]$ \\  
\hline 
$\C_{lu}$&   $[-0.49,  0.19]$     & $\C_{lq}^{(3)}$ & $[ -0.32,    0.57]$   & $\C_{lq}^{(1)}$ & $[ -0.25,  0.66]$\\ 
\hline
$\C_{le}$ & $[ -5.3, 11]$      & $\C_{ld}$ & $[ -3.8,   8.7]$    & $\C_{dd}$& $[ -51,  26]$  \\
\hline 
$\C_{ed}$ & $[ -12, 6.7]$    & $\C_{ee}$ & $[  -3.9,   2.4]$    & $\C_{eu}$ & $[ -0.36,  0.58]$\\    
\hline
$\C_{ud}^{(1)}$ &  $  [-3.0,         5.6   ]$ &&&&\\
\hline\hline
\end{tabular}
\caption{$95\%$ confidence level allowed ranges for single parameter fit to coefficients not  contributing to the
lowest order predictions. The scale $\Lambda$ is taken to be $1~TeV$.  \label{tab:nlopred}}
\end{center}
\end{table}

\begin{table}
\begin{center}
\begin{tabular}{|c|c|c|}
\hline
Coefficient& LO & NLO\nonumber \\ 
\hline\hline
$\C_{\phi D}$ & $[-0.034,0.041]$   & [-0.039,0.051]\\
\hline
$\C_{\phi WB}$&  $[-0.080, 0.0021]$ & $[-0.098,  0.012]$\\
\hline
$\C_{\phi  d}$ & $ [-0.81, -0.093]$ &  $[-1.07,-0.03]$\\
\hline
$\C_{\phi l}^{(3)}$ &  $ [-0.025, 0.12] $  & $ [-0.039, 0.16]$\\
\hline
$\C_{\phi u}$  &  $ [-0.12, 0.37]$ & $[-0.21, 0.41]$\\
\hline 
$\C_{\phi  l}^{(1)}$ &  $[-0.0086, 0.036]$ &  $[-0.0072, 0.037]$\\
\hline
$\C_{ll}$ &  $[-0.085, 0.035]$ &  $[-0.088,  0.033]$\\
\hline
$\C_{\phi  q}^{(1)}$ & $[-0.060, 0.076]$ &  $[-0.095, 0.076]$\\
\hline\hline
\end{tabular}
\caption{$95\%$ confidence level allowed ranges for fit to coefficients marginalizing over the  other 7 operators 
we are considering.  The coefficients of all  operators not listed in the table are set to 0.
The scale $\Lambda$ is taken to be $1~TeV$.  \label{tab:marg}}
\end{center}
\end{table}

  At 
lowest order, the $\chi^2_{LO}$ is sensitive to $8$ combinations of operators, implying that there are $2$ blind 
directions\cite{Falkowski:2014tna,Corbett:2012ja,Elias-Miro:2013mua}.  These $8$ combinations can be thought of as
the combinations of operators contributing to $\delta g_{L}^{Zu}, \delta g_{L}^{Zd}, \delta g_{L}^{Ze},\delta g_{L}^{Z\nu},
\delta g_{R}^{Zu}, \delta g_{R}^{Zd}, \delta g_{R}^{Ze}$, and $M_W$.  Because of the $SU(2)$ symmetry of Eq. \ref{eq:su2rel},
at LO there is no additional information from $\Gamma_W$.   
Since our study includes only $14$ data points, we clearly cannot fit to all of  the SMEFT coefficients appearing at one loop
At NLO, the fit is sensitive to  only $10$ combinations of operators.  The
additional information can be thought of as coming from $\delta g_L^{Zb}$ and $\Gamma_W$ where the top quark makes significant
contributions.  Since there are $32$ coefficients that contribute to the NLO fit to the electroweak observables, resolving these $22$ blind directions 
requires input from other processes and/or assumptions about which operators can be safely neglected.  

We chose to perform our  fits setting $\C_{\phi e}=0$ and 
$\C_{\phi q}^{(3)}=0$, along with setting all of the operators that first appear at NLO to $0$.  We then marginalize over the remaining
operators to study the numerical impacts of the NLO contributions.  These results are shown in 
Tab. \ref{tab:marg}.  We see that the effects of the NLO corrections can be significant, although the numerical results are sensitive
to which operators are set to $0$.  Our results suggest that including the NLO corrections in the global fits (where the complete set of
operators can potentially be bounded) may be important.  

\begin{figure}
  \centering
\includegraphics[width=0.48\textwidth]{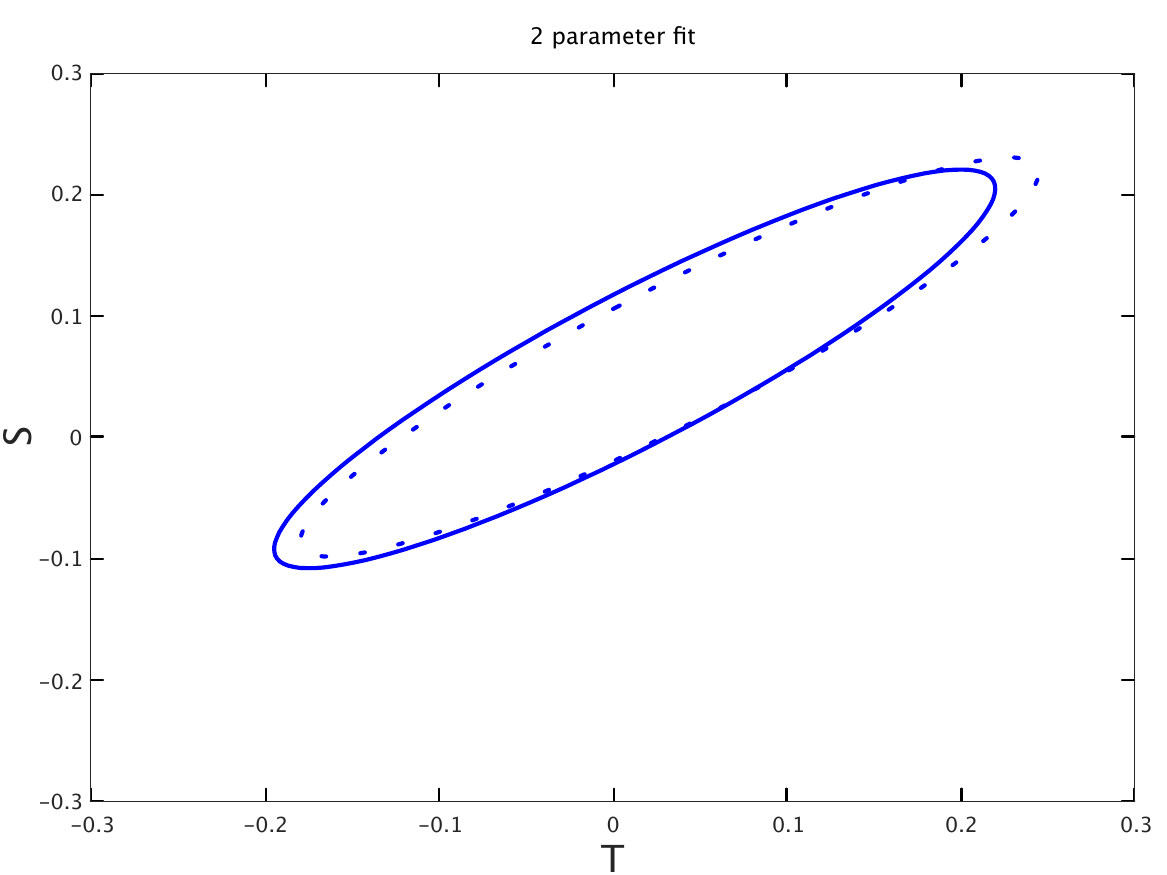}
 \caption{$95\%$ CL limits from a $2$ parameter fit to $\C_{\phi WB}$ and $\C_{\phi D}$, setting all other coefficients to $0$.  The
 scale $\Lambda=1~TeV$.The solid line is the result of the LO fit, while the dotted line is the NLO fit to the electroweak parameters of this study. }
   \label{fig:st}
\end{figure}
As another way of examining the impact of the NLO contributions, we consider the oblique parameters.   The tree level
SMEFT  contributions are,
\begin{eqnarray}
\alpha \Delta S&=& 4 c_W s_W{v^2\over \Lambda^2}C_{\phi W B}
\nonumber \\
\alpha \Delta T&=& -{v^2\over 2 \Lambda^2}\C_{\phi D}\, .
\end{eqnarray}
 For the NLO oblique parameter fit, we set all coefficients
 to $0$,  except $\C_{\phi W B} $ and $\C_{\phi D}$. 
 The resulting limits are shown in Fig. \ref{fig:st}, (where what we
are really plotting are the limits on the coefficients from a $2$ parameter fit to our observables).  In this example, the
effect of the NLO SMEFT corrections is small.  At NLO, new coefficients can influence
the oblique parameters, and the complete one-loop SMEFT result is given in Ref. \cite{Chen:2013kfa}.

 Our calculation includes only the resonant $Z$ and $W$ contributions to the precision electroweak observables.
  In the SMEFT, there are tree level
non-resonant contributions due to $4$-fermion operators that  can complicate the experimental extraction of the widths from
the data.  The size of these effects has been estimated in Ref. \cite{Berthier:2015oma}.  For example,
\begin{equation}
\delta \Gamma(Z\rightarrow {\text{hadrons})}\sim 0.6~\mev \C_{4f}\biggl({1~TeV\over \Lambda}\biggr)^2\, ,
\end{equation}
where $\C_{4f}$ is a generic $4$-fermion operator.  These effects could potentially be similar in size to the electroweak
corrections we have computed.   There are also off-shell corrections proportional to ${q^2\over \Lambda^2}$, where $q^2$
is the momentum running through the $Z$ boson propagator.  The experimental cuts \cite{ALEPH:2005ab} were designed to extract predominantly the
on-shell $Z$ events, so although the size of these effects is expected to be small a detailed theoretical study would be needed
to quantify them.

\section{Conclusions}
\label{sec:conc}
We have computed the NLO electroweak and QCD corrections to the SMEFT predictions for the precision electroweak
observables.  Our results are presented in a numerical form that can easily be incorporated in the global fitting programs. 
We also present numerical results for the LO and NLO $\chi^2$ that can be customized for the reader's use.  Our studies
suggest that the NLO SMEFT corrections may have a sizable effect on the global fits. 
Numerical results for the SMEFT NLO expressions for the observables considered here, along with the $\chi^2_{LO}$, $\chi^2_{NLO}$ and the 
matrix $M_{NLO}$, are posted at
\url{https://quark.phy.bnl.gov/Digital_Data_Archive/dawson/ewpo_19}.
 \section*{Acknowledgements}
We thank Sam Homiller
for discussions.  V4:  We thank Anke Biekoetter, Ben Pecjak, Darren Scott,
and   Tommy Smith for a detailed  and extremely useful numerical comparison of results.  Revised results (with very minor changes) are posted in the digital data file.
S.D.   is  supported by the U.S. Department of Energy under Grant Contract  de-sc0012704.  
P.P.G. is supported by the Spanish Research Agency (Agencia Estatal de Investigacion) through the contract FPA2016-78022-P and IFT Centro de Excelencia Severo Ochoa un- der grant SEV-2016-0597.

\appendix
\section{Observables to LO and NLO in the SMEFT }
\label{sec:appresults}
In this appendix, we report the contributions to the observables of Table \ref{tab:ferm} using
the definitions of Eq. \ref{eq:defs}. The contributions to the $Z$ width are,
\begin{eqnarray}
\delta \Gamma(Z\rightarrow \nu {\overline \nu})^{LO}&=& \vLam\biggl\{
-0.3318 \Tfla + 0.1659 \Tll - 0.0829 \TphiD
 \biggr\}\, \gev
 \nonumber \\
\delta \Gamma(Z\rightarrow \nu {\overline \nu})^{NLO}&=& \vLam\biggl\{
- 0.3446 \Tfla+ 0.1640 \Tll - 0.0853 \TphiD 
-0.0003 \Tfd - 0.0003 \Tfe  \nonumber \\ &&- 0.0018 \Tflc - 0.0073 \Tfqa + 
 0.0054 \Tfqc + 0.0083 \Tfu - 0.0004 \Tld \nonumber \\ &&- 0.0004 \Tle - 
 0.0061 \Tlqa - 0.0061 \Tlqc + 0.008 \Tlu  - 
 0.0002 \Tphik - 0.0001 \TphiW \nonumber \\ &&+ 0.0063 \TphiWB + 0.0001 \TuW - 0.0001 \TW
 \biggr\}\, \gev
  \nonumber \\
\delta  \Gamma(Z\rightarrow l^+l^-)^{LO}&=& \vLam\biggl\{
-0.1408 \Tfe + 0.191 \Tfla - 0.037 \Tflc + 0.114 \Tll - 0.057 \TphiD  \nonumber \\ &&- 
 0.0713 \TphiWB
 \biggr\}\, \gev
\nonumber \\ 
\delta  \Gamma(Z\rightarrow l^+l^-)^{NLO}&=& \vLam\biggl\{
- 0.1596 \Tfe + 0.1834 \Tfla - 0.0221 \Tflc  + 0.0985 \Tll - 0.0508 \TphiD \nonumber \\ &&
- 
 0.0349 \TphiWB  - 0.0001 \TphiW
-0.0002 \Ted - 0.0005 \Tee + 0.0035 \Teu
\nonumber \\ &&
 - 0.0002 \Tfd  - 0.0042 \Tfqa + 0.0032 \Tfqc + 0.0049 \Tfu + 
 0.0002 \Tld 
 \nonumber \\ &&
 + 0.0001 \Tle + 0.0034 \Tlqa - 0.0031 \Tlqc - 
 0.0045 \Tlu  - 0.0001 \Tphik  
 \nonumber \\ &&- 0.0027 \Tqe - 0.0007 \TuB - 0.0007 \TuW - 0.0001 \TW
 \biggr\}\, \gev
\nonumber 
\end{eqnarray}
\begin{eqnarray}
 \delta \Gamma(Z\rightarrow u {\overline u})^{LO}&=& \delta \Gamma(Z\rightarrow c {\overline c})^{LO}
  \nonumber \\
  &=&
\vLam\biggl\{
-0.9261 \Tflc - 0.7138 \Tfqa + 0.7138 \Tfqc + 0.2815 \Tfu + 0.4631 \Tll \nonumber \\ && - 
 0.2315 \TphiD - 0.4093 \TphiWB
 \biggr\}\, \gev
  \nonumber \\
  \delta \Gamma(Z\rightarrow u {\overline u})^{NLO}&=& \delta \Gamma(Z\rightarrow c {\overline c})^{NLO}
  \nonumber \\
&=&\vLam\biggl\{
- 0.9620 \Tflc -  0.7559 \Tfqa + 0.7477 \Tfqc + 0.3515 \Tfu + 0.4634 \Tll \nonumber \\ &&
- 0.2402 \TphiD - 0.4016 \TphiWB 
+0.0004 \Teu - 0.0013 \Tfd - 0.0013 \Tfe \nonumber \\ && - 0.0022 \Tfla  - 0.0009 \Tlqa + 
 0.0099 \Tlqc + 0.0004 \Tlu - 0.0002 \TphiB  \nonumber \\ &&- 
 0.0003 \Tphik - 0.0004 \TphiW - 0.0009 \Tqda - 
 0.0009 \Tqe - 0.0448 \Tqqa \nonumber \\ &&- 0.0629 \Tqqc + 0.0223 \Tqua - 0.006 \TuB + 
 0.0004 \Tuda - 0.0221 \Tuu \nonumber \\ &&- 0.0049 \TuW - 0.0005 \TW
 \biggr\}\, \gev
  \nonumber \\
\delta \Gamma(Z\rightarrow d {\overline d})^{LO}&=&\delta \Gamma(Z\rightarrow s {\overline s})^{LO}\nonumber \\
&=& \vLam\biggl\{
-0.1408 \Tfd - 1.0299 \Tflc + 0.8545 \Tfqa + 0.8545 \Tfqc + 0.5149 \Tll  \nonumber \\ && - 
 0.2575 \TphiD - 0.3379 \TphiWB
 \biggr\}\, \gev
  \nonumber \\
  \delta \Gamma(Z\rightarrow d {\overline d})^{NLO}&=&\delta \Gamma(Z\rightarrow s {\overline s})^{NLO}\nonumber \\
  &=& \vLam\biggl\{
  - 0.1659  \Tfd - 1.1057 \Tflc + 0.8818 \Tfqa + 0.9150 \Tfqc + 0.5329 \Tll \nonumber \\ &&
  -  0.2757 \TphiD - 0.3582  \TphiWB
  -0.0004 \Tdd - 0.0002 \Ted  - 0.0013 \Tfe \nonumber \\ && - 0.0024 \Tfla  + 0.0255 \Tfu - 0.0002 \Tld  + 0.0011 \Tlqa + 0.011 \Tlqc \nonumber \\ && - 0.0002 \TphiB - 0.0004 \Tphik - 0.0004 \TphiW  - 0.0016 \Tqda \nonumber \\ && + 0.0011 \Tqe + 0.0293 \Tqqa - 0.0118 \Tqqc - 0.0205 \Tqua \nonumber \\ && - 0.0053 \TuB + 0.0035 \Tuda - 0.0041 \TuW - 0.0005 \TW
 \biggr\}\, \gev
 \nonumber\\
\delta \Gamma(Z\rightarrow b {\overline b})^{LO}&=&\vLam\biggl\{
-0.1400 \Tfd - 1.0242 \Tflc + 0.8498 \Tfqa + 0.8498 \Tfqc + 0.5121 \Tll \nonumber \\ &&- 
 0.2561 \TphiD - 0.3361 \TphiWB
 \biggr\}\, \gev
\nonumber 
\end{eqnarray}
\begin{eqnarray}
\delta \Gamma(Z\rightarrow b {\overline b})^{NLO}&=&\vLam\biggl\{
- 0.1649 \Tfd - 1.0666 \Tflc + 0.8724 \Tfqa + 0.8790 \Tfqc +0.5134 \Tll \nonumber \\ &&- 
 0.2659  \TphiD- 0.3413  \TphiWB
-0.0004 \Tdd - 0.0002 \Ted  - 0.0013 \Tfe \nonumber \\ && - 0.0023 \Tfla  + 0.0222 \Tfu - 0.0002 \Tld + 0.0011 \Tlqa + 0.0109 \Tlqc \nonumber \\ && - 0.0002 \TphiB  - 0.0004 \Tphik - 0.0004 \TphiW  - 
 0.0016 \Tqda + 0.0011 \Tqe \nonumber \\ && + 0.0292 \Tqqa - 0.0117 \Tqqc - 0.0204 \Tqua - 
 0.0069 \TuB + 0.0035 \Tuda \nonumber \\ && - 0.0168 \TuW - 0.0027\TW
 \biggr\}\, \gev\, .
 \nonumber
 \end{eqnarray}
 The SMEFT contributions to the total $Z$ width are, 
\begin{eqnarray}
\delta \Gamma_Z^{LO}&=&\vLam\biggl\{
-0.4223 \Tfd - 0.4223 \Tfe - 0.4223 \Tfla - 5.053 \Tflc + 1.1361 \Tfqa \nonumber \\ && + 
 3.9911 \Tfqc + 0.5631 \Tfu + 3.3106 \Tll - 1.6553 \TphiD - 2.0463 \TphiWB
 \biggr\}\, \gev
\nonumber \\ 
\delta \Gamma_Z^{NLO}&=&\vLam\biggl\{
 - 0.5008 \Tfd - 0.4862 \Tfe - 0.4951 \Tfla - 5.2739 \Tflc + 1.0898 \Tfqa \nonumber \\ && 
 + 4.2302 \Tfqc +  0.8157 \Tfu + 3.2934 \Tll- 1.7061  \TphiD - 1.9465 \TphiWB \nonumber \\ && 
-0.0013 \Tdd - 0.0011 \Ted - 0.0016 \Tee + 0.0113 \Teu- 0.0011 \Tld - 0.0011 \Tle\nonumber \\ &&  - 0.0065 \Tlqa + 
 0.025 \Tlqc + 0.0113 \Tlu - 0.0014 \TphiB  - 
 0.0028 \Tphik - 0.0026 \TphiW  \nonumber \\ &&- 0.0065 \Tqda - 
 0.0065 \Tqe - 0.0017 \Tqqa - 0.161 \Tqqc - 0.0168 \Tqua - 0.0318 \TuB\nonumber \\ && + 
 0.0113 \Tuda - 0.0443 \Tuu - 0.0365 \TuW - 0.0054 \TW
 \biggr\}\, \gev\, .
\nonumber 
\end{eqnarray}

The ratios are defined to be
\begin{eqnarray} 
R_l & =& {\Sigma_q\Gamma(Z\rightarrow q {\overline q})\over \Gamma(Z\rightarrow ll)}\nonumber \\
R_c& =&{\Gamma(Z\rightarrow u {\overline {u}})\over \Sigma_q\Gamma(Z\rightarrow q {\overline q}) }\nonumber \\ 
R_b & =&{ \Gamma(Z\rightarrow d {\overline d)}\over 
\Sigma_q\Gamma(Z\rightarrow q {\overline q)}}\, ,
\nonumber 
\end{eqnarray}
and the SMEFT contributions are,
\begin{eqnarray}
\delta R_l^{LO}&=&\vLam\biggl\{-4.978 \Tfd + 33.673 \Tfe - 45.688 \Tfla - 49.393 \Tflc + 13.39 \Tfqa + 
 47.041 \Tfqc \nonumber \\ &&  + 6.637 \Tfu + 1.853 \Tll - 0.926 \TphiD - 4.532 \TphiWB
 \biggr\}
\nonumber \\
\delta R_l^{NLO}&=&\vLam\biggl\{
 -5.8831 \Tfd + 39.214 \Tfe -45.510 \Tfla -56.368 \Tflc +14.398 \Tfqa +49.304 \Tfqc \nonumber \\ && + 
 8.0458 \Tfu + 5.3901 \Tll - 2.8457 \TphiD - 13.262 \TphiWB 
-0.015 \Tdd + 0.038 \Ted + 0.124 \Tee \nonumber \\ && - 0.834 \Teu -0.063 \Tld -0.012 \Tle  - 0.795 \Tlqa + 
1.362 \Tlqc + 1.083 \Tlu - 0.004 \TphiB \nonumber \\ && - 0.002 \Tphik - 
0.005 \TphiW - 0.077 \Tqda + 0.654 \Tqe - 0.020 \Tqqa -
1.898 \Tqqc - 0.198 \Tqua \nonumber \\ && - 0.168 \TuB + 0.133 \Tuda - 0.522 \Tuu - 
 0.254 \TuW - 0.037 \TW
 \biggr\}
\nonumber \\
\delta R_c^{LO}&=&\vLam\biggl\{
0.0421 \Tfd - 0.0449 \Tflc - 0.5279 \Tfqa + 0.0164 \Tfqc + 0.1073 \Tfu \nonumber \\ && + 
 0.0224 \Tll - 0.0112 \TphiD - 0.0549 \TphiWB
 \biggr\}
\nonumber \\
\delta R_c^{NLO}&=&\vLam\biggl\{
 0.0487 \Tfd - 0.0380 \Tflc - 0.5455 \Tfqa + 0.0136 \Tfqc + 0.1253 \Tfu \nonumber \\ &&
+ 0.0182 \Tll - 0.0096 \TphiD - 0.0465 \TphiWB
+0.0001 \Tdd + 0.0001 \Ted + 0.0001 \Teu  \nonumber \\ && - 0.0001 \Tfe - 
 0.0001 \Tfla  + 
 0.0001 \Tld  - 0.0007 \Tlqa + 0.0005 \Tlqc + 0.0001 \Tlu  \nonumber \\ && + 0.0001 \Tqda - 0.0007 \Tqe - 
 0.0258 \Tqqa - 0.0205 \Tqqc + 0.0146 \Tqua - 0.0005 \TuB \nonumber \\ && - 0.0009 \Tuda - 
 0.0084 \Tuu + 0.0006 \TuW + 0.0002 \TW
 \biggr\}
\nonumber \\
\delta R_b^{LO}&=&\vLam\biggl\{
-0.02808 \Tfd + 0.02993 \Tflc + 0.3519 \Tfqa - 0.01094 \Tfqc \nonumber \\ && - 
 0.07156 \Tfu - 0.01497 \Tll + 0.00748 \TphiD + 0.03661 \TphiWB
 \biggr\}
\nonumber \\
\delta R_b^{NLO}&=&\vLam\biggl\{
 - 0.03276 \Tfd + 0.03271 \Tflc + 0.3618 \Tfqa - 
 0.01678 \Tfqc \nonumber \\ && - 0.08398 \Tfu - 0.01583 \Tll + 
 0.00828  \TphiD + 0.03465 \TphiWB
-0.00008 \Tdd  \nonumber \\ &&- 0.00004 \Ted - 0.00009 \Teu + 
 0.00007 \Tfe + 0.00010 \Tfla - 0.00004 \Tld \nonumber \\ && + 
 0.00044 \Tlqa - 0.00034 \Tlqc - 0.00009 \Tlu + 0.00001 \TphiB + 0.00001 \TphiW \nonumber \\ && - 0.00008 \Tqda + 
 0.00044 \Tqe + 0.01717 \Tqqa + 0.01367 \Tqqc - 0.00972 \Tqua \nonumber \\ && - 
 0.00028 \TuB + 0.00060 \Tuda + 0.00562 \Tuu - 0.00534 \TuW - 0.00098 \TW
 \biggr\}\, .
 \nonumber 
\end{eqnarray}

 \begin{eqnarray}
\sigma_h^{LO}&=&\vLam\biggl\{
4.05 \Tfd - 55.524 \Tfe + 109.235 \Tfla + 32.796 \Tflc - 10.896 \Tfqa \nonumber \\ && - 
 38.278 \Tfqc - 5.4 \Tfu + 4.319 \Tll - 2.16 \TphiD - 10.565 \TphiWB
\biggr\}\, {\rm nb}
\nonumber \\
\sigma_h^{NLO}&=&\vLam\biggl\{
 4.656 \Tfd - 
 62.835 \Tfe + 106.805 \Tfla + 40.603 \Tflc - 11.515  \Tfqa \nonumber \\ && - 
 38.995 \Tfqc - 6.267 \Tfu - 1.176 \Tll + 0.738 \TphiD + 3.178 \TphiWB
+0.012 \Tdd \nonumber \\ &&- 0.068 \Ted - 0.205 \Tee + 1.382 \Teu + 0.142 \Tld + 0.063 \Tle + 
 1.945 \Tlqa - 1.101 \Tlqc \nonumber \\ &&- 2.598 \Tlu - 0.001 \TphiB + 
 0.002 \Tphik + 0.063 \Tqda - 1.064 \Tqe\nonumber \\ && + 
 0.017 \Tqqa + 1.544 \Tqqc + 0.161 \Tqua - 0.009 \TuB - 0.108 \Tuda + 
 0.424 \Tuu\nonumber \\ && + 0.067 \TuW + 0.027 \TW
\biggr\}\, {\rm nb}
\nonumber
\end{eqnarray}

The asymmetries are defined as,
\begin{eqnarray}
A_l&=& {\Gamma(Z\rightarrow e^+_Le^-_L)-\Gamma(Z\rightarrow e^+_Re^-_R)\over
\Gamma(Z\rightarrow e^+ e^-)}\nonumber \\
A_c&=&
{\Gamma(Z\rightarrow u_L{\overline u}_L)-\Gamma(Z\rightarrow u_R {\overline u}_R)\over
\Gamma(Z\rightarrow u {\overline u})}
\nonumber \\
A_b &=&{\Gamma(Z\rightarrow d_L{\overline d}_L)-\Gamma(Z\rightarrow d_R {\overline d}_R)\over
\Gamma(Z\rightarrow d {\overline d})}\, ,
\nonumber
\end{eqnarray}
and the SMEFT contributions are, 
\begin{eqnarray}
\delta A_l^{LO}&=&\vLam\biggl\{
2.1503 \Tfe + 1.5848 \Tfla - 2.7529 \Tflc + 2.1689 \Tll - 1.0844 \TphiD - 
 5.305 \TphiWB
\biggr\}
\nonumber \\
\delta A_l^{NLO}&=&\vLam\biggl\{
 + 2.1666 \Tfe + 
 1.8745 \Tfla - 3.3587 \Tflc + 2.5342 \Tll - 1.3250 \TphiD - 6.1599 \TphiWB \nonumber \\ &&
+0.0027 \Ted + 0.0076 \Tee - 0.0518 \Teu - 0.0128 \Tfd - 0.0867 \Tfqa + 0.0719 \Tfqc\nonumber \\ && + 0.1190 \Tfu + 
 0.0021 \Tld + 0.0049 \Tle + 0.0307 \Tlqa + 0.0098 \Tlqc - 
 0.0406 \Tlu\nonumber \\ && - 0.0026 \TphiB - 0.0004 \Tphik - 
 0.0026 \TphiW + 0.0392 \Tqe\nonumber \\ && - 0.0866 \TuB - 
 0.0711 \TuW - 0.0046 \TW
\biggr\}
\nonumber \\
\delta A_c^{LO}&=&\vLam\biggl\{
-1.779 \Tflc - 0.65 \Tfqa + 0.65 \Tfqc - 1.648 \Tfu + 0.889 \Tll - 
 0.445 \TphiD - 2.175 \TphiWB
\biggr\}
\nonumber \\
\delta A_c^{NLO}&=&\vLam\biggl\{
 - 2.295 \Tflc - 
 0.867 \Tfqa + 0.856 \Tfqc - 1.798 \Tfu + 1.110 \Tll - 0.581 \TphiD - 
 2.699 \TphiWB \nonumber \\ &&
-0.002 \Teu - 0.006 \Tfd - 0.006 \Tfe - 0.007 \Tfla - 0.001 \Tlqa + 
 0.025 \Tlqc - 0.002 \Tlu \nonumber \\ && - 0.001 \TphiB - 0.001 \TphiW - 0.001 \Tqda - 0.001 \Tqe - 0.045 \Tqqa - 0.063 \Tqqc - 
 0.014 \Tqua \nonumber \\ && - 0.037 \TuB - 0.002 \Tuda + 0.13 \Tuu - 0.03 \TuW - 0.002 \TW
\biggr\}
\nonumber \\
\delta A_b^{LO}&=&\vLam\biggl\{
0.727 \Tfd - 0.328 \Tflc + 0.12 \Tfqa + 0.12 \Tfqc + 0.164 \Tll - 
 0.082 \TphiD - 0.401 \TphiWB
\biggr\}
\nonumber \\
\delta A_b^{NLO}&=&\vLam\biggl\{
 + 0.842 \Tfd - 
 0.424 \Tflc + 0.149 \Tfqa + 0.157 \Tfqc + 
 0.205 \Tll - 0.107 \TphiD - 0.501 \TphiWB \nonumber \\ &&
+0.002 \Tdd + 0.001 \Ted - 0.001 \Tfe - 0.001 \Tfla + 0.009 \Tfu + 0.001 \Tld + 0.005 \Tlqc \nonumber \\ && + 0.014 \Tqda + 
 0.005 \Tqqa - 0.002 \Tqqc - 0.003 \Tqua \nonumber \\ && - 0.007 \TuB - 0.018 \Tuda - 
 0.007 \TuW - 0.001 \TW\, .
\biggr\} \nonumber 
\end{eqnarray}

Finally, the forward backward asymmetries are defined as 
\begin{equation}
A_{FB,i}={\sigma_F-\sigma_B\over \sigma_F+\sigma_B}\, ,
\end{equation}
where defining $\theta$ to be the angle between the incoming $l^-$ and the outgoing ${\overline f}_i$,
$\sigma_F$ has $\theta$ between $(0,{\pi\over 2})$ and $\sigma_B$ has $\theta$ between $({\pi\over 2},\pi)$.
The SMEFT results are, 
\begin{eqnarray}
A_{FB,l}^{LO}&=&\vLam\biggl\{
0.9547 \Tfe + 0.7037 \Tfla - 1.2223 \Tflc + 0.9630 \Tll - 0.4815 \TphiD - 
 2.3555 \TphiWB
\biggr\}
\nonumber \\
A_{FB,l}^{NLO}&=&\vLam\biggl\{
 + 0.4783 \Tfe + 
 0.4138 \Tfla - 0.7414 \Tflc + 0.5594 \Tll - 0.2925 \TphiD - 1.3598 \TphiWB \nonumber \\ &&
+0.0006 \Ted + 0.0017 \Tee - 0.0114 \Teu - 0.0028 \Tfd - 0.0191 \Tfqa + 0.0159 \Tfqc + 0.0263 \Tfu \nonumber \\ && + 
 0.0005 \Tld + 0.0011 \Tle + 0.0068 \Tlqa + 0.0022 \Tlqc - 
 0.009 \Tlu - 0.0006 \TphiB - 0.0001 \Tphik \nonumber \\ && - 
 0.0006 \TphiW + 0.0086 \Tqe - 0.0191 \TuB - 
 0.0157 \TuW - 0.0010 \TW
\biggr\}
\nonumber \\
A_{FB,c}^{LO}&=&\vLam\biggl\{
1.1785 \Tfe + 0.8686 \Tfla - 1.9036 \Tflc - 0.1443 \Tfqa + 0.1443 \Tfqc  \nonumber \\ && - 
 0.3658 \Tfu + 1.3861 \Tll - 0.693 \TphiD - 3.3903 \TphiWB
\biggr\}
\nonumber \\
A_{FB,c}^{NLO}&=&\vLam\biggl\{
 + 1.0846 \Tfe + 
 0.9381 \Tfla - 1.9356 \Tflc - 0.1391 \Tfqa + 0.1305 \Tfqc  \nonumber \\ && - 0.1388 \Tfu + 1.3918 \Tll - 0.7278 \TphiD - 3.3833 \TphiWB
+0.0014 \Ted + 0.0038 \Tee  \nonumber \\ && - 0.0262 \Teu - 0.007 \Tfd + 
 0.0011 \Tld + 0.0024 \Tle + 0.0153 \Tlqa + 0.0076 \Tlqc  \nonumber \\ && - 
 0.0206 \Tlu - 0.0014 \TphiB - 0.0002 \Tphik - 
 0.0014 \TphiW - 0.0001 \Tqda + 0.0195 \Tqe  \nonumber \\ && - 
 0.0050 \Tqqa - 0.0070 \Tqqc - 0.0016 \Tqua - 0.0475 \TuB  \nonumber \\ && - 0.0002 \Tuda + 
 0.0143 \Tuu - 0.0389 \TuW - 0.0025 \TW
\biggr\}
\nonumber \\
A_{FB,b}^{LO}&=&\vLam\biggl\{
0.1615 \Tfd + 1.5275 \Tfe + 1.1258 \Tfla - 2.0284 \Tflc + 0.0266 \Tfqa  \nonumber \\ && + 
 0.0266 \Tfqc + 1.5771 \Tll - 0.7886 \TphiD - 3.8576 \TphiWB
\biggr\}
\nonumber \\
A_{FB,b}^{NLO}&=&\vLam\biggl\{
 + 0.0840 \Tfd + 
  1.5062 \Tfe + 1.3031 \Tfla - 2.3819 \Tflc - 0.0439 \Tfqa  \nonumber \\ && + 
  0.0673 \Tfqc + 1.7845 \Tll - 
  0.9331 \TphiD - 4.3379 \TphiWB
+0.0002 \Tdd  \nonumber \\ && + 0.0020 \Ted + 0.0053 \Tee - 0.0360 \Teu + 0.0838 \Tfu + 0.0016 \Tld + 0.0034 \Tle  \nonumber \\ && + 
  0.0214 \Tlqa + 0.0073 \Tlqc - 0.0283 \Tlu - 0.0018 \TphiB - 0.0003 \Tphik - 0.0018 \TphiW  \nonumber \\ && + 
  0.0015 \Tqda + 0.0273 \Tqe + 0.0005 \Tqqa - 0.0002 \Tqqc - 
  0.0003 \Tqua  \nonumber \\ && - 0.0610 \TuB - 0.0020 \Tuda - 0.0502 \TuW - 0.0033 \TW
\biggr\} \, .
\nonumber 
\end{eqnarray}

\bibliographystyle{utphys}
\bibliography{mw}

\providecommand{\href}[2]{#2}\begingroup\raggedright\begin{thebibliography}{10}

\bibitem{Dawson:2018dcd}
S.~Dawson, C.~Englert, and T.~Plehn, ``{Higgs Physics: It ain't over till it's
  over},'' \href{http://dx.doi.org/10.1016/j.physrep.2019.05.001}{{\em Phys.
  Rept.} {\bfseries 816} (2019) 1--85},
\href{http://arxiv.org/abs/1808.01324}{{\ttfamily arXiv:1808.01324 [hep-ph]}}.

\bibitem{Brivio:2017vri}
I.~Brivio and M.~Trott, ``{The Standard Model as an Effective Field Theory},''
  \href{http://dx.doi.org/10.1016/j.physrep.2018.11.002}{{\em Phys. Rept.}
  {\bfseries 793} (2019) 1--98},
\href{http://arxiv.org/abs/1706.08945}{{\ttfamily arXiv:1706.08945 [hep-ph]}}.

\bibitem{Biekotter:2018rhp}
A.~Biekotter, T.~Corbett, and T.~Plehn, ``{The Gauge-Higgs Legacy of the LHC
  Run II},'' \href{http://dx.doi.org/10.21468/SciPostPhys.6.6.064}{{\em SciPost
  Phys.} {\bfseries 6} (2019) 064},
\href{http://arxiv.org/abs/1812.07587}{{\ttfamily arXiv:1812.07587 [hep-ph]}}.

\bibitem{Almeida:2018cld}
E.~da~Silva~Almeida, A.~Alves, N.~Rosa~Agostinho, O.~J.~P. Eboli, and M.~C.
  Gonzalez-Garcia, ``{Electroweak Sector Under Scrutiny: A Combined Analysis of
  LHC and Electroweak Precision Data},''
  \href{http://dx.doi.org/10.1103/PhysRevD.99.033001}{{\em Phys. Rev.}
  {\bfseries D99} no.~3, (2019) 033001},
\href{http://arxiv.org/abs/1812.01009}{{\ttfamily arXiv:1812.01009 [hep-ph]}}.

\bibitem{deBlas:2017wmn}
J.~de~Blas, M.~Ciuchini, E.~Franco, S.~Mishima, M.~Pierini, L.~Reina, and
  L.~Silvestrini, ``{The Global Electroweak and Higgs Fits in the LHC era},''
  \href{http://dx.doi.org/10.22323/1.314.0467}{{\em PoS} {\bfseries
  EPS-HEP2017} (2017) 467},
\href{http://arxiv.org/abs/1710.05402}{{\ttfamily arXiv:1710.05402 [hep-ph]}}.

\bibitem{DiVita:2017eyz}
S.~Di~Vita, C.~Grojean, G.~Panico, M.~Riembau, and T.~Vantalon, ``{A global
  view on the Higgs self-coupling},''
  \href{http://dx.doi.org/10.1007/JHEP09(2017)069}{{\em JHEP} {\bfseries 09}
  (2017) 069},
\href{http://arxiv.org/abs/1704.01953}{{\ttfamily arXiv:1704.01953 [hep-ph]}}.

\bibitem{Ellis:2018gqa}
J.~Ellis, C.~W. Murphy, V.~Sanz, and T.~You, ``{Updated Global SMEFT Fit to
  Higgs, Diboson and Electroweak Data},''
  \href{http://dx.doi.org/10.1007/JHEP06(2018)146}{{\em JHEP} {\bfseries 06}
  (2018) 146},
\href{http://arxiv.org/abs/1803.03252}{{\ttfamily arXiv:1803.03252 [hep-ph]}}.

\bibitem{Grojean:2018dqj}
C.~Grojean, M.~Montull, and M.~Riembau, ``{Diboson at the LHC vs LEP},''
  \href{http://dx.doi.org/10.1007/JHEP03(2019)020}{{\em JHEP} {\bfseries 03}
  (2019) 020},
\href{http://arxiv.org/abs/1810.05149}{{\ttfamily arXiv:1810.05149 [hep-ph]}}.

\bibitem{Berthier:2016tkq}
L.~Berthier, M.~Bjorn, and M.~Trott, ``{Incorporating doubly resonant $W^\pm$
  data in a global fit of SMEFT parameters to lift flat directions},''
  \href{http://dx.doi.org/10.1007/JHEP09(2016)157}{{\em JHEP} {\bfseries 09}
  (2016) 157},
\href{http://arxiv.org/abs/1606.06693}{{\ttfamily arXiv:1606.06693 [hep-ph]}}.

\bibitem{Pomarol:2013zra}
A.~Pomarol and F.~Riva, ``{Towards the Ultimate SM Fit to Close in on Higgs
  Physics},'' \href{http://dx.doi.org/10.1007/JHEP01(2014)151}{{\em JHEP}
  {\bfseries 01} (2014) 151},
\href{http://arxiv.org/abs/1308.2803}{{\ttfamily arXiv:1308.2803 [hep-ph]}}.

\bibitem{Baglio:2017bfe}
J.~Baglio, S.~Dawson, and I.~M. Lewis, ``{An NLO QCD effective field theory
  analysis of $W^+W^-$ production at the LHC including fermionic operators},''
  \href{http://dx.doi.org/10.1103/PhysRevD.96.073003}{{\em Phys. Rev.}
  {\bfseries D96} no.~7, (2017) 073003},
\href{http://arxiv.org/abs/1708.03332}{{\ttfamily arXiv:1708.03332 [hep-ph]}}.

\bibitem{Baglio:2018bkm}
J.~Baglio, S.~Dawson, and I.~M. Lewis, ``{NLO effects in EFT fits to $W^+W^-$
  production at the LHC},''
  \href{http://dx.doi.org/10.1103/PhysRevD.99.035029}{{\em Phys. Rev.}
  {\bfseries D99} no.~3, (2019) 035029},
\href{http://arxiv.org/abs/1812.00214}{{\ttfamily arXiv:1812.00214 [hep-ph]}}.

\bibitem{Hartland:2019bjb}
N.~P. Hartland, F.~Maltoni, E.~R. Nocera, J.~Rojo, E.~Slade, E.~Vryonidou, and
  C.~Zhang, ``{A Monte Carlo global analysis of the Standard Model Effective
  Field Theory: the top quark sector},''
  \href{http://dx.doi.org/10.1007/JHEP04(2019)100}{{\em JHEP} {\bfseries 04}
  (2019) 100},
\href{http://arxiv.org/abs/1901.05965}{{\ttfamily arXiv:1901.05965 [hep-ph]}}.

\bibitem{BuarqueFranzosi:2017jrj}
D.~Buarque~Franzosi, E.~Vryonidou, and C.~Zhang, ``{Scalar production and decay
  to top quarks including interference effects at NLO in QCD in an EFT
  approach},'' \href{http://dx.doi.org/10.1007/JHEP10(2017)096}{{\em JHEP}
  {\bfseries 10} (2017) 096},
\href{http://arxiv.org/abs/1707.06760}{{\ttfamily arXiv:1707.06760 [hep-ph]}}.

\bibitem{Durieux:2018ggn}
G.~Durieux, J.~Gu, E.~Vryonidou, and C.~Zhang, ``{Probing top-quark couplings
  indirectly at Higgs factories},''
  \href{http://dx.doi.org/10.1088/1674-1137/42/12/123107}{{\em Chin. Phys.}
  {\bfseries C42} no.~12, (2018) 123107},
\href{http://arxiv.org/abs/1809.03520}{{\ttfamily arXiv:1809.03520 [hep-ph]}}.

\bibitem{Boughezal:2019xpp}
R.~Boughezal, C.-Y. Chen, F.~Petriello, and D.~Wiegand, ``{Top quark decay at
  next-to-leading order in the Standard Model Effective Field Theory},''
\href{http://arxiv.org/abs/1907.00997}{{\ttfamily arXiv:1907.00997 [hep-ph]}}.

\bibitem{Cullen:2019nnr}
J.~M. Cullen, B.~D. Pecjak, and D.~J. Scott, ``{NLO corrections to $h\to b\bar
  b$ decay in SMEFT},''
\href{http://arxiv.org/abs/1904.06358}{{\ttfamily arXiv:1904.06358 [hep-ph]}}.

\bibitem{Gauld:2016kuu}
R.~Gauld, B.~D. Pecjak, and D.~J. Scott, ``{QCD radiative corrections for $h\to
  b\bar b$ in the Standard Model Dimension-6 EFT},''
  \href{http://dx.doi.org/10.1103/PhysRevD.94.074045}{{\em Phys. Rev.}
  {\bfseries D94} no.~7, (2016) 074045},
\href{http://arxiv.org/abs/1607.06354}{{\ttfamily arXiv:1607.06354 [hep-ph]}}.

\bibitem{Gauld:2015lmb}
R.~Gauld, B.~D. Pecjak, and D.~J. Scott, ``{One-loop corrections to $h\to b\bar
  b$ and $h\to \tau\bar \tau$ decays in the Standard Model Dimension-6 EFT:
  four-fermion operators and the large-$m_t$ limit},''
  \href{http://dx.doi.org/10.1007/JHEP05(2016)080}{{\em JHEP} {\bfseries 05}
  (2016) 080},
\href{http://arxiv.org/abs/1512.02508}{{\ttfamily arXiv:1512.02508 [hep-ph]}}.

\bibitem{Dawson:2018liq}
S.~Dawson and P.~P. Giardino, ``{Electroweak corrections to Higgs boson decays
  to $\gamma\gamma$ and $W^+W^-$ in standard model EFT},''
  \href{http://dx.doi.org/10.1103/PhysRevD.98.095005}{{\em Phys. Rev.}
  {\bfseries D98} no.~9, (2018) 095005},
\href{http://arxiv.org/abs/1807.11504}{{\ttfamily arXiv:1807.11504 [hep-ph]}}.

\bibitem{Dedes:2018seb}
A.~Dedes, M.~Paraskevas, J.~Rosiek, K.~Suxho, and L.~Trifyllis, ``{The decay
  $h\to \gamma\gamma$ in the Standard-Model Effective Field Theory},''
  \href{http://dx.doi.org/10.1007/JHEP08(2018)103}{{\em JHEP} {\bfseries 08}
  (2018) 103},
\href{http://arxiv.org/abs/1805.00302}{{\ttfamily arXiv:1805.00302 [hep-ph]}}.

\bibitem{Hartmann:2015aia}
C.~Hartmann and M.~Trott, ``{Higgs Decay to Two Photons at One Loop in the
  Standard Model Effective Field Theory},''
  \href{http://dx.doi.org/10.1103/PhysRevLett.115.191801}{{\em Phys. Rev.
  Lett.} {\bfseries 115} no.~19, (2015) 191801},
\href{http://arxiv.org/abs/1507.03568}{{\ttfamily arXiv:1507.03568 [hep-ph]}}.

\bibitem{Hartmann:2015oia}
C.~Hartmann and M.~Trott, ``{On one-loop corrections in the standard model
  effective field theory; the $\Gamma(h \rightarrow \gamma \, \gamma)$ case},''
  \href{http://dx.doi.org/10.1007/JHEP07(2015)151}{{\em JHEP} {\bfseries 07}
  (2015) 151},
\href{http://arxiv.org/abs/1505.02646}{{\ttfamily arXiv:1505.02646 [hep-ph]}}.

\bibitem{Dawson:2018pyl}
S.~Dawson and P.~P. Giardino, ``{Higgs decays to $ZZ$ and $Z\gamma$ in the
  standard model effective field theory: An NLO analysis},''
  \href{http://dx.doi.org/10.1103/PhysRevD.97.093003}{{\em Phys. Rev.}
  {\bfseries D97} no.~9, (2018) 093003},
\href{http://arxiv.org/abs/1801.01136}{{\ttfamily arXiv:1801.01136 [hep-ph]}}.

\bibitem{Dedes:2019bew}
A.~Dedes, K.~Suxho, and L.~Trifyllis, ``{The decay $h\to Z \gamma$ in the
  Standard-Model Effective Field Theory},''
  \href{http://dx.doi.org/10.1007/JHEP06(2019)115}{{\em JHEP} {\bfseries 06}
  (2019) 115},
\href{http://arxiv.org/abs/1903.12046}{{\ttfamily arXiv:1903.12046 [hep-ph]}}.

\bibitem{Dawson:2019xfp}
S.~Dawson, P.~P. Giardino, and A.~Ismail, ``{Standard model EFT and the
  Drell-Yan process at high energy},''
  \href{http://dx.doi.org/10.1103/PhysRevD.99.035044}{{\em Phys. Rev.}
  {\bfseries D99} no.~3, (2019) 035044},
\href{http://arxiv.org/abs/1811.12260}{{\ttfamily arXiv:1811.12260 [hep-ph]}}.

\bibitem{Dawson:2018jlg}
S.~Dawson and A.~Ismail, ``{Standard model EFT corrections to Z boson
  decays},'' \href{http://dx.doi.org/10.1103/PhysRevD.98.093003}{{\em Phys.
  Rev.} {\bfseries D98} no.~9, (2018) 093003},
\href{http://arxiv.org/abs/1808.05948}{{\ttfamily arXiv:1808.05948 [hep-ph]}}.

\bibitem{Trott:2017yhn}
M.~Trott, ``{EWPD in the SMEFT and the $\mathcal{O}(y_t^2,\lambda)$ one loop
  $Z$ decay width},'' in {\em {Proceedings, 52nd Rencontres de Moriond on
  Electroweak Interactions and Unified Theories: La Thuile, Italy, March 18-25,
  2017}}, pp.~63--70.
\newblock 2017.
\newblock
\href{http://arxiv.org/abs/1705.05652}{{\ttfamily arXiv:1705.05652 [hep-ph]}}.
\newblock

\bibitem{Hartmann:2016pil}
C.~Hartmann, W.~Shepherd, and M.~Trott, ``{The $Z$ decay width in the SMEFT:
  $y_t$ and $\lambda$ corrections at one loop},''
  \href{http://dx.doi.org/10.1007/JHEP03(2017)060}{{\em JHEP} {\bfseries 03}
  (2017) 060},
\href{http://arxiv.org/abs/1611.09879}{{\ttfamily arXiv:1611.09879 [hep-ph]}}.

\bibitem{Buchmuller:1985jz}
W.~Buchmuller and D.~Wyler, ``{Effective Lagrangian Analysis of New
  Interactions and Flavor Conservation},''
\href{http://dx.doi.org/10.1016/0550-3213(86)90262-2}{{\em Nucl. Phys.}
  {\bfseries B268} (1986) 621--653}.

\bibitem{Grzadkowski:2010es}
B.~Grzadkowski, M.~Iskrzynski, M.~Misiak, and J.~Rosiek, ``{Dimension-Six Terms
  in the Standard Model Lagrangian},''
  \href{http://dx.doi.org/10.1007/JHEP10(2010)085}{{\em JHEP} {\bfseries 10}
  (2010) 085},
\href{http://arxiv.org/abs/1008.4884}{{\ttfamily arXiv:1008.4884 [hep-ph]}}.

\bibitem{Dedes:2017zog}
A.~Dedes, W.~Materkowska, M.~Paraskevas, J.~Rosiek, and K.~Suxho, ``{Feynman
  rules for the Standard Model Effective Field Theory in R$_{\xi}$ -gauges},''
  \href{http://dx.doi.org/10.1007/JHEP06(2017)143}{{\em JHEP} {\bfseries 06}
  (2017) 143},
\href{http://arxiv.org/abs/1704.03888}{{\ttfamily arXiv:1704.03888 [hep-ph]}}.

\bibitem{Alonso:2013hga}
R.~Alonso, E.~E. Jenkins, A.~V. Manohar, and M.~Trott, ``{Renormalization Group
  Evolution of the Standard Model Dimension Six Operators III: Gauge Coupling
  Dependence and Phenomenology},''
  \href{http://dx.doi.org/10.1007/JHEP04(2014)159}{{\em JHEP} {\bfseries 04}
  (2014) 159},
\href{http://arxiv.org/abs/1312.2014}{{\ttfamily arXiv:1312.2014 [hep-ph]}}.

\bibitem{Berthier:2015oma}
L.~Berthier and M.~Trott, ``{Towards consistent Electroweak Precision Data
  constraints in the SMEFT},''
  \href{http://dx.doi.org/10.1007/JHEP05(2015)024}{{\em JHEP} {\bfseries 05}
  (2015) 024},
\href{http://arxiv.org/abs/1502.02570}{{\ttfamily arXiv:1502.02570 [hep-ph]}}.

\bibitem{Hollik:1988ii}
W.~F.~L. Hollik, ``{Radiative Corrections in the Standard Model and their Role
  for Precision Tests of the Electroweak Theory},''
\href{http://dx.doi.org/10.1002/prop.2190380302}{{\em Fortsch. Phys.}
  {\bfseries 38} (1990) 165--260}.

\bibitem{Dubovyk:2019szj}
I.~Dubovyk, A.~Freitas, J.~Gluza, T.~Riemann, and J.~Usovitsch, ``{Electroweak
  pseudo-observables and Z-boson form factors at two-loop accuracy},''
\href{http://arxiv.org/abs/1906.08815}{{\ttfamily arXiv:1906.08815 [hep-ph]}}.

\bibitem{Jenkins:2013zja}
E.~E. Jenkins, A.~V. Manohar, and M.~Trott, ``{Renormalization Group Evolution
  of the Standard Model Dimension Six Operators I: Formalism and lambda
  Dependence},'' \href{http://dx.doi.org/10.1007/JHEP10(2013)087}{{\em JHEP}
  {\bfseries 10} (2013) 087},
\href{http://arxiv.org/abs/1308.2627}{{\ttfamily arXiv:1308.2627 [hep-ph]}}.

\bibitem{Jenkins:2013wua}
E.~E. Jenkins, A.~V. Manohar, and M.~Trott, ``{Renormalization Group Evolution
  of the Standard Model Dimension Six Operators II: Yukawa Dependence},''
  \href{http://dx.doi.org/10.1007/JHEP01(2014)035}{{\em JHEP} {\bfseries 01}
  (2014) 035},
\href{http://arxiv.org/abs/1310.4838}{{\ttfamily arXiv:1310.4838 [hep-ph]}}.

\bibitem{Chen:2013kfa}
C.-Y. Chen, S.~Dawson, and C.~Zhang, ``{Electroweak Effective Operators and
  Higgs Physics},'' \href{http://dx.doi.org/10.1103/PhysRevD.89.015016}{{\em
  Phys. Rev.} {\bfseries D89} no.~1, (2014) 015016},
\href{http://arxiv.org/abs/1311.3107}{{\ttfamily arXiv:1311.3107 [hep-ph]}}.

\bibitem{Ghezzi:2015vva}
M.~Ghezzi, R.~Gomez-Ambrosio, G.~Passarino, and S.~Uccirati, ``{NLO Higgs
  effective field theory and $\kappa$-framework},''
  \href{http://dx.doi.org/10.1007/JHEP07(2015)175}{{\em JHEP} {\bfseries 07}
  (2015) 175},
\href{http://arxiv.org/abs/1505.03706}{{\ttfamily arXiv:1505.03706 [hep-ph]}}.

\bibitem{Hahn:2000kx}
T.~Hahn, ``{Generating Feynman diagrams and amplitudes with FeynArts 3},''
  \href{http://dx.doi.org/10.1016/S0010-4655(01)00290-9}{{\em Comput. Phys.
  Commun.} {\bfseries 140} (2001) 418--431},
\href{http://arxiv.org/abs/hep-ph/0012260}{{\ttfamily arXiv:hep-ph/0012260
  [hep-ph]}}.

\bibitem{Alloul:2013bka}
A.~Alloul, N.~D. Christensen, C.~Degrande, C.~Duhr, and B.~Fuks, ``{FeynRules
  2.0 - A complete toolbox for tree-level phenomenology},''
  \href{http://dx.doi.org/10.1016/j.cpc.2014.04.012}{{\em Comput. Phys.
  Commun.} {\bfseries 185} (2014) 2250--2300},
\href{http://arxiv.org/abs/1310.1921}{{\ttfamily arXiv:1310.1921 [hep-ph]}}.

\bibitem{Mertig:1990an}
R.~Mertig, M.~Bohm, and A.~Denner, ``{FEYN CALC: Computer algebraic calculation
  of Feynman amplitudes},''
\href{http://dx.doi.org/10.1016/0010-4655(91)90130-D}{{\em Comput. Phys.
  Commun.} {\bfseries 64} (1991) 345--359}.

\bibitem{Shtabovenko:2016sxi}
V.~Shtabovenko, R.~Mertig, and F.~Orellana, ``{New Developments in FeynCalc
  9.0},'' \href{http://dx.doi.org/10.1016/j.cpc.2016.06.008}{{\em Comput. Phys.
  Commun.} {\bfseries 207} (2016) 432--444},
\href{http://arxiv.org/abs/1601.01167}{{\ttfamily arXiv:1601.01167 [hep-ph]}}.

\bibitem{Hahn:2000jm}
T.~Hahn, ``{Automatic loop calculations with FeynArts, FormCalc, and
  LoopTools},'' \href{http://dx.doi.org/10.1016/S0920-5632(00)00848-3}{{\em
  Nucl. Phys. Proc. Suppl.} {\bfseries 89} (2000) 231--236},
\href{http://arxiv.org/abs/hep-ph/0005029}{{\ttfamily arXiv:hep-ph/0005029
  [hep-ph]}}.

\bibitem{Anastasiou:2002yz}
C.~Anastasiou and K.~Melnikov, ``{Higgs boson production at hadron colliders in
  NNLO QCD},'' \href{http://dx.doi.org/10.1016/S0550-3213(02)00837-4}{{\em
  Nucl. Phys.} {\bfseries B646} (2002) 220--256},
  \href{http://arxiv.org/abs/hep-ph/0207004}{{\ttfamily arXiv:hep-ph/0207004
  [hep-ph]}}.

\bibitem{Tarasov:1997kx}
O.~V. Tarasov, ``{Generalized recurrence relations for two loop propagator
  integrals with arbitrary masses},''
  \href{http://dx.doi.org/10.1016/S0550-3213(97)00376-3}{{\em Nucl. Phys.}
  {\bfseries B502} (1997) 455--482},
\href{http://arxiv.org/abs/hep-ph/9703319}{{\ttfamily arXiv:hep-ph/9703319
  [hep-ph]}}.

\bibitem{Martin:2003qz}
S.~P. Martin, ``{Evaluation of two loop selfenergy basis integrals using
  differential equations},''
  \href{http://dx.doi.org/10.1103/PhysRevD.68.075002}{{\em Phys. Rev.}
  {\bfseries D68} (2003) 075002},
\href{http://arxiv.org/abs/hep-ph/0307101}{{\ttfamily arXiv:hep-ph/0307101
  [hep-ph]}}.

\bibitem{Smirnov:2014hma}
A.~V. Smirnov, ``{FIRE5: a C++ implementation of Feynman Integral REduction},''
  \href{http://dx.doi.org/10.1016/j.cpc.2014.11.024}{{\em Comput. Phys.
  Commun.} {\bfseries 189} (2015) 182--191},
\href{http://arxiv.org/abs/1408.2372}{{\ttfamily arXiv:1408.2372 [hep-ph]}}.

\bibitem{Freitas:2014hra}
A.~Freitas, ``{Higher-order electroweak corrections to the partial widths and
  branching ratios of the Z boson},''
  \href{http://dx.doi.org/10.1007/JHEP04(2014)070}{{\em JHEP} {\bfseries 04}
  (2014) 070},
\href{http://arxiv.org/abs/1401.2447}{{\ttfamily arXiv:1401.2447 [hep-ph]}}.

\bibitem{Dubovyk:2018rlg}
I.~Dubovyk, A.~Freitas, J.~Gluza, T.~Riemann, and J.~Usovitsch, ``{Complete
  electroweak two-loop corrections to Z boson production and decay},''
  \href{http://dx.doi.org/10.1016/j.physletb.2018.06.037}{{\em Phys. Lett.}
  {\bfseries B783} (2018) 86--94},
\href{http://arxiv.org/abs/1804.10236}{{\ttfamily arXiv:1804.10236 [hep-ph]}}.

\bibitem{ALEPH:2005ab}
{\bfseries ALEPH, DELPHI, L3, OPAL, SLD, LEP Electroweak Working Group, SLD
  Electroweak Group, SLD Heavy Flavour Group} Collaboration, S.~Schael {\em
  et~al.}, ``{Precision electroweak measurements on the $Z$ resonance},''
  \href{http://dx.doi.org/10.1016/j.physrep.2005.12.006}{{\em Phys. Rept.}
  {\bfseries 427} (2006) 257--454},
\href{http://arxiv.org/abs/hep-ex/0509008}{{\ttfamily arXiv:hep-ex/0509008
  [hep-ex]}}.

\bibitem{Awramik:2006uz}
M.~Awramik, M.~Czakon, and A.~Freitas, ``{Electroweak two-loop corrections to
  the effective weak mixing angle},''
  \href{http://dx.doi.org/10.1088/1126-6708/2006/11/048}{{\em JHEP} {\bfseries
  11} (2006) 048},
\href{http://arxiv.org/abs/hep-ph/0608099}{{\ttfamily arXiv:hep-ph/0608099
  [hep-ph]}}.

\bibitem{Awramik:2008gi}
M.~Awramik, M.~Czakon, A.~Freitas, and B.~A. Kniehl, ``{Two-loop electroweak
  fermionic corrections to sin**2 theta**b anti-b(eff)},''
  \href{http://dx.doi.org/10.1016/j.nuclphysb.2008.12.031}{{\em Nucl. Phys.}
  {\bfseries B813} (2009) 174--187},
\href{http://arxiv.org/abs/0811.1364}{{\ttfamily arXiv:0811.1364 [hep-ph]}}.

\bibitem{CDF:2016cry}
{\bfseries CDF, D0} Collaboration, T.~T. E.~W. Group, ``{Combination of the CDF
  and D0 Effective Leptonic Electroweak Mixing Angles},''
\newblock 2016.
\newblock
\url{http://lss.fnal.gov/archive/2016/conf/fermilab-conf-16-295-e.pdf}.
\newblock

\bibitem{PhysRevD.98.030001}
{\bfseries Particle Data Group} Collaboration, ``Review of particle physics,''
  \href{http://dx.doi.org/10.1103/PhysRevD.98.030001}{{\em Phys. Rev. D}
  {\bfseries 98} (Aug, 2018) 030001}.
  \url{https://link.aps.org/doi/10.1103/PhysRevD.98.030001}.

\bibitem{Awramik:2003rn}
M.~Awramik, M.~Czakon, A.~Freitas, and G.~Weiglein, ``{Precise prediction for
  the W boson mass in the standard model},''
  \href{http://dx.doi.org/10.1103/PhysRevD.69.053006}{{\em Phys. Rev.}
  {\bfseries D69} (2004) 053006},
\href{http://arxiv.org/abs/hep-ph/0311148}{{\ttfamily arXiv:hep-ph/0311148
  [hep-ph]}}.

\bibitem{Erler:2019hds}
J.~Erler and M.~Schott, ``{Electroweak Precision Tests of the Standard Model
  after the Discovery of the Higgs Boson},''
  \href{http://dx.doi.org/10.1016/j.ppnp.2019.02.007}{{\em Prog. Part. Nucl.
  Phys.} {\bfseries 106} (2019) 68--119},
\href{http://arxiv.org/abs/1902.05142}{{\ttfamily arXiv:1902.05142 [hep-ph]}}.

\bibitem{Cho:2011rk}
G.-C. Cho, K.~Hagiwara, Y.~Matsumoto, and D.~Nomura, ``{The MSSM confronts the
  precision electroweak data and the muon g-2},''
  \href{http://dx.doi.org/10.1007/JHEP11(2011)068}{{\em JHEP} {\bfseries 11}
  (2011) 068},
\href{http://arxiv.org/abs/1104.1769}{{\ttfamily arXiv:1104.1769 [hep-ph]}}.

\bibitem{Berthier:2015gja}
L.~Berthier and M.~Trott, ``{Consistent constraints on the Standard Model
  Effective Field Theory},''
  \href{http://dx.doi.org/10.1007/JHEP02(2016)069}{{\em JHEP} {\bfseries 02}
  (2016) 069},
\href{http://arxiv.org/abs/1508.05060}{{\ttfamily arXiv:1508.05060 [hep-ph]}}.

\bibitem{Corbett:2017qgl}
T.~Corbett, O.~J.~P. Eboli, and M.~C. Gonzalez-Garcia, ``{Unitarity Constraints
  on Dimension-six Operators II: Including Fermionic Operators},''
  \href{http://dx.doi.org/10.1103/PhysRevD.96.035006}{{\em Phys. Rev.}
  {\bfseries D96} no.~3, (2017) 035006},
\href{http://arxiv.org/abs/1705.09294}{{\ttfamily arXiv:1705.09294 [hep-ph]}}.

\bibitem{Falkowski:2014tna}
A.~Falkowski and F.~Riva, ``{Model-independent precision constraints on
  dimension-6 operators},''
  \href{http://dx.doi.org/10.1007/JHEP02(2015)039}{{\em JHEP} {\bfseries 02}
  (2015) 039},
\href{http://arxiv.org/abs/1411.0669}{{\ttfamily arXiv:1411.0669 [hep-ph]}}.

\bibitem{Corbett:2012ja}
T.~Corbett, O.~J.~P. Eboli, J.~Gonzalez-Fraile, and M.~C. Gonzalez-Garcia,
  ``{Robust Determination of the Higgs Couplings: Power to the Data},''
  \href{http://dx.doi.org/10.1103/PhysRevD.87.015022}{{\em Phys. Rev.}
  {\bfseries D87} (2013) 015022},
\href{http://arxiv.org/abs/1211.4580}{{\ttfamily arXiv:1211.4580 [hep-ph]}}.

\bibitem{Elias-Miro:2013mua}
J.~Elias-Miro, J.~R. Espinosa, E.~Masso, and A.~Pomarol, ``{Higgs windows to
  new physics through d=6 operators: constraints and one-loop anomalous
  dimensions},'' \href{http://dx.doi.org/10.1007/JHEP11(2013)066}{{\em JHEP}
  {\bfseries 11} (2013) 066},
\href{http://arxiv.org/abs/1308.1879}{{\ttfamily arXiv:1308.1879 [hep-ph]}}.

\end{thebibliography}\endgroup


\providecommand{\href}[2]{#2}\begingroup\raggedright\endgroup

\end{document}